\documentclass[lineno]{jfm}

\usepackage{graphicx}
\usepackage{multirow}
\usepackage{newtxtext}
\usepackage{newtxmath}
\usepackage{natbib}
\usepackage{hyperref}
\hypersetup{
    colorlinks = true,
    urlcolor   = blue,
    citecolor  = black,
}
\newcommand{\RomanNumeralCaps}[1]
\linenumbers
\usepackage{longtable}

\title{Effect of radius ratio on the sheared annular centrifugal turbulent convection}

\author{Jun Zhong\aff{1},
  Junyi Li\aff{1}\corresp{\email{junyili@mail.tsinghua.edu.cn}}
 \and Chao Sun\aff{1,2}
 \corresp{\email{chaosun@tsinghua.edu.cn}}}

\affiliation{\aff{1} New Cornerstone Science Laboratory, Center for Combustion Energy, Key Laboratory for
Thermal Science and Power Engineering of Ministry of Education, Department of Energy and
Power Engineering, Tsinghua University, 100084 Beijing, China
\aff{2} Department of Engineering Mechanics, School of Aerospace Engineering, Tsinghua University, 100084 Beijing, China}

\begin{document}
\maketitle

\begin{abstract}
We perform the linear stability analysis and direct numerical simulations to study the effect of radius ratio on the instability and flow characteristics of the sheared annular centrifugal Rayleigh-Bénard convection (ACRBC), where the cold inner cylinder and the hot outer cylinder rotate with a small angular velocity difference. With the shear enhancement, the thermal convection is suppressed and finally gets stable for different radius ratios $\eta\in[0.2, 0.95]$. Considering the inhomogeneous distribution of shear stresses in the base flow, a new global Richardson number $Ri_g$ is defined and the marginal-state curves for different radius ratios are successfully unified in the parameter domain of $Ri_g$ and the Rayleigh number $Ra$. The results are consistent with the marginal-state curve of the wall-sheared classical RBC in the streamwise direction, demonstrating that the basic stabilization mechanisms are identical. Moreover, systems with small radius ratios exhibit greater geometric asymmetry. On the one hand, this results in a smaller equivalent aspect ratio for the system, accommodating fewer convection roll pairs. Fewer roll pairs are more likely to cause a transition in the flow structure during shear enhancement. On the other hand, the shear distribution is more inhomogeneous, allowing for an outward shift of the convection region and the elevation of bulk temperature under strong shear. 
\end{abstract}

%\clearpage
%{\bf MSC Codes }  {\it(Optional)} Please enter your MSC Codes here

\section{Introduction}

Thermally driven turbulent flows are ubiquitous in nature and industrial processes. As a general paradigm for modeling this common phenomenon, the Rayleigh-Bénard convection (RBC) has been studied extensively in scientific research \citep{ahlers_heat_2009,lohse_small-scale_2010,chilla_new_2012,xia_current_2013,ecke_2023_turbulent}, in which a layer of fluid is confined between two horizontal plates, heated from below and cooled from above. Under gravity or other body force fields, buoyancy is generated, inducing instability, driving thermal convection, and forming manifold and involute flow structures \citep{niemela_wind_2001,xi_laminar_2004,sun_three-dimensional_2005,wang_heat_2021,guo2023flow}. In recent years, apart from the classical RBC with rectangular cells, annular centrifugal Rayleigh-Bénard convection (ACRBC) has been put forward \citep{jiang_supergravitational_2020, wang_effects_2022, wang_statistics_2023}. Due to the use of stronger centrifugal force to substitute gravity, a higher Rayleigh number can be achieved in ACRBC, enhancing the thermal convection to the ultimate regime \citep{jiang_experimental_2022}. The scaling law in ACRBC is found in agreement with the theoretical predictions \citep{grossmann_scaling_2000, grossmann_multiple_2011}. Similar to RBC, Taylor-Coutte (TC) flow, where the flow is impelled by two concentric cylinders rotating independently with constant angular velocity, is another canonical paradigm of the physics of fluids to model the flow driven by wall shear stress \citep{Huisman_Statistics_2013, grossmann_highreynolds_2016}. In the TC flow, differential angular speed induces instabilities and forms the secondary flow including Taylor rolls. As similar exact global balance relations between the respective driving and the dissipation can be derived, a close analogy is put forward between RBC and TC flow, by which the Grossmann–Lohse theory is extended from RBC to TC flow \citep{bradshaw_analogy_1969, Eckhardt_scaling_2000, eckhardt_torque_2007,busse_twins_2012}. 
%because for both RBC and TC systems, similar exact global balance relations between the respective driving and the dissipation can be derived.

%In the Taylor-Coutte (TC) system, the flow is impelled by two concentric cylinders rotating independently with constant angular velocity \citep{Eckhardt_scaling_2000, eckhardt_torque_2007, Huisman_Statistics_2013, grossmann_highreynolds_2016}. Differential angular speed induces instabilities and forms the secondary flow including Taylor rolls. There is a close analogy between RB and TC flow, as for both systems exact global balance relations between the respective driving and the dissipation can be derived \citep{bradshaw_analogy_1969, eckhardt_torque_2007, grossmann_highreynolds_2016}. 

The comprehensive study of the interplay between buoyancy and shear holds significant importance in enhancing our comprehension of atmospheric motion and oceanic flow. \citep{Dearorff_Numerical_1972, Khanna_three_1998, Egbers2014, Feng_Season_2022}. Numerous attempts have been made to integrate shear and buoyancy within a unified system with the intent of investigating their mutual coupling effects, including wall-sheared RBC \citep{deardorff_gravitational_1965, blass_flow_2020, blass_effect_2021} and TC system with axial or radial temperature difference under gravity or centrifugal force \citep{yoshikawa_instability_2013, meyer_effect_2015, kang_radial_2017, leng_flow_2021, leng_mutual_2022}. Recently, based on the high similarity between ACRBC and TC systems, we have proposed an innovative system, namely the sheared ACRBC system, combining ACRBC with TC to study the coupling effect of shear and buoyancy (see \citet{zhong_sheared_2023}, ZWS23 for short). In the new system, an ACBRC cell bounded by two independent-rotating concentric cylinders is considered. It is a closed system and inherits the exact global balance relations from ACRBC and TC. The system becomes ACRBC when the two cylinders rotate at the same angular velocity and turns into TC flow when two cylinders rotate at different speeds with no temperature difference. In the large parameter domain of buoyancy strength and shear strength, it is found that an ACRBC flow gets stable at first and then develops into a TC flow with the enhancement of shear. In such a system with a fixed geometry, the coupling mechanism of buoyancy and shear is well revealed.

To further reveal the coupling mechanism of buoyancy and shear in a sheared ACRBC system, it is necessary to consider the effect of the radius ratio. In the TC flow, with the radius ratio increasing, the momentum Nusselt number is found to increase at first and then saturate \citep{grossmann_highreynolds_2016}. In the ACRBC system, the radius ratio has a significant impact on the critical Rayleigh number of the convection onset, heat transfer efficiency, bulk temperature, and zonal flow \citep{pitz_onset_2017, wang_effects_2022}. Meanwhile, the study on the radius ratio is a key to link the sheared ACRBC system to the wall-sheared RBC, as these two systems may gradually become identical when the radius ratio tends to one. Therefore, in this paper, we concentrate on the radius ratio effect, attempting to give a more complete and systematic understanding of the coupling effect of buoyancy and shear in the sheared ACRBC.

The rest of the paper is organized as follows: the governing equations are introduced in section \ref{sec:geq}, and the results of linear stability analysis (LSA) and direct numerical simulation (DNS) are demonstrated in section \ref{sec:lsa} and \ref{sec:dns}, respectively. Finally, conclusions are presented in section \ref{sec:con}.

\section{Governing equations}\label{sec:geq}

\begin{figure}
    \centering
    \includegraphics[width=0.9\linewidth]{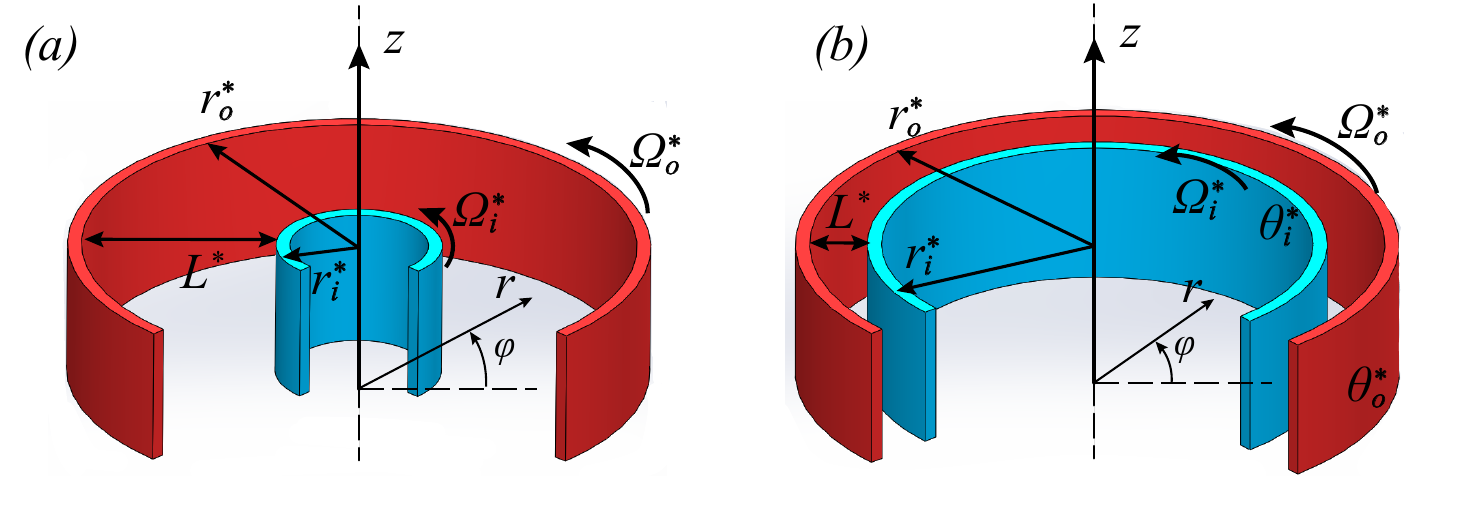}
    \captionsetup{justification=raggedright}
    \caption{Schematic diagram of the flow configuration in the sheared ACRBC system with {\textit{(a)}} a small radius ratio $\eta=r^*_i/r^*_o=0.3$ and {\textit{(b)}} a large radius ratio $\eta=0.8$ in the stationary reference frame. $r^*_{i,o}, \Omega^*_{i,o}$ and $\theta^*_{i,o}$ are the radius, angular speed, and temperature of the inner and outer cylinders, respectively. $L^*$ is the gap between two cylinders.}
	\label{fig: 1}
\end{figure}

In sheared ACRBC, incompressible viscous fluid is bounded by an inner cylinder with radius $r_i^*$ and an outer cylinder with radius $r_o^*$, rotating independently about $z$ axis. Hereafter, the asterisk $*$ denotes the dimensional variables. The radius ratio is then defined as $\eta=r^*_i/r^*_o$. Figure \ref{fig: 1} depicts two typical flow domains with  $\eta=0.3$ and $0.8$. The inner cold cylinder with temperature $\theta_i^*$ rotates at a larger angular velocity $\Omega_i^*$, while the outer hot cylinder rotates at a smaller angular velocity $\Omega_o^*$. $L^*=r^*_o-r^*_i$ is the gap width and $\Delta^*=\theta^*_{o}-\theta^*_{i}$ is the temperature difference between the two cylinders. No-slip and isothermal boundary conditions are applied at two cylinder surfaces, and periodic boundary conditions are imposed on the velocity and temperature in the axial direction. In the rotating frame with averaged angular velocity $\Omega_c^*=(\Omega^*_i+\Omega^*_o)/2$, an equivalent gravitational acceleration along the radial direction can be defined as $g_e={\Omega_c^*}^2(r^*_i+r^*_o)/2$. Then the free fall velocity $U^*=\sqrt{ g_e\alpha\Delta^* L^*}$, the gap $L^*$ and the temperature difference $\Delta^*$ are introduced as velocity, length, and temperature scales, respectively. The coefficient of thermal expansion $\alpha$, the kinematic viscosity $\nu$, and the thermal diffusivity $\kappa$ of the fluid are assumed to be constant. Then the motion of the flow is governed by the non-dimensional Oberbeck–Boussinesq equation, which reads \citep{jiang_supergravitational_2020, zhong_sheared_2023}:

%A three-dimensional annular centrifugal RB cell bounded by two independent-rotating concentric cylinders is considered, as shown in figure \ref{fig: 1}. The inner cylinder with radius $R_i$ rotates about $z$ axis at angular velocity $\Omega_i$ and the outer cylinder with radius $R_o$ rotates at the angular velocity $\Omega_o$. $L=R_o-R_i$ is the gap width between the two cylinders. The temperature difference between the hot outer cylinder and the cold inner cylinder is $\Delta=\theta_{o}-\theta_{i}$. For the boundary, no-slip and isothermal conditions are applied at two cylinder surfaces; periodic boundary conditions are imposed on the velocity and temperature in the axial direction. In the rotating frame with angular velocity $\Omega_c=(\Omega_i+\Omega_o)/2$, the motion of the flow is governed by the non-dimensional Oberbeck–Boussinesq equation \citep{jiang_supergravitational_2020, zhong_sheared_2023}:

\begin{equation}\label{eq: OB}
    \begin{aligned}
    \boldsymbol{\nabla}\cdot\boldsymbol{u}&=0,\\
    \frac{\partial\boldsymbol{u}}{\partial t}+\boldsymbol{u}\cdot\boldsymbol{\nabla u}=-\boldsymbol{\nabla} p-Ro^{-1}\boldsymbol{e_z}\times\boldsymbol{u}&+\sqrt{\frac{Pr}{Ra}}\nabla^2\boldsymbol{u}-\theta\frac{2(1-\eta)}{1+\eta}(1+\frac{2u_\varphi}{Ro^{-1}r})^2\boldsymbol{r},\\
    \frac{\partial \theta}{\partial t}+\boldsymbol{\nabla}\cdot (\boldsymbol{u}\theta)&=\sqrt{\frac{1}{Ra\cdot Pr}}\nabla^2 \theta,\\
    \end{aligned}
\end{equation}
where $\boldsymbol{u}=(u_r,u_\varphi,u_z)$ is the velocity vector, $p$ is the pressure, $\theta$ is the temperature,  $\boldsymbol{e_z}$ is the unit vector in the axial direction and $\eta=r^*_i/r^*_o$ is the radius ratio. 
%Scaled quantities, including $L=R_o-R_i$ for length, $\Delta$ for temperature, $U=\sqrt{\alpha\Delta \Omega_c^2\frac{(R_i+R_o)}{2}L}$ for velocity, and $L/U$ for time are used to non-dimensionalize the governing equation, where $\alpha$ is the coefficient of thermal expansion of the fluid. 
Relative to $\Omega^*_c$, the non-dimensional boundary conditions read:
\begin{equation}\label{eq: BC}
\begin{aligned}
r&=r_i: \boldsymbol{u}=(0,\Omega ,0), \theta=0,\\
r&=r_o: \boldsymbol{u}=(0,-\Omega ,0), \theta=1,
\end{aligned}
\end{equation}
where $r_i=\eta/(1-\eta)$ and $r_o=1/(1-\eta)$ are the non-dimensional radii of the inner and outer cylinders, and $\Omega=(\Omega^*_i-\Omega^*_c)L^*/U^*$ represents the non-dimensional rotating angular velocity difference.

The above dimensionless governing equations and the boundary conditions reveal five control parameters in the current system: the Rayleigh number $Ra$, the inverse Rossby number $Ro^{-1}$, the Prandtl number $Pr$, the angular velocity difference $\Omega$ and the radius ratio $\eta$, in which $Ra$, $Ro$, and $Pr$ are defined as:

%Three non-dimensional parameters are generated through non-dimensionalization in the governing equations, they are the Rayleigh number $Ra$ (measuring the buoyancy-driving strength), the inverse Rossby number $Ro^{-1}$ (measuring Coriolis effects), and the Prandtl number $Pr$ (fluid property), expressed as:
\begin{equation}\label{eq: Par}       
        Ra=\frac{g_e\alpha\Delta^*{L^*}^3}{\nu\kappa}, \quad Ro^{-1}=\frac{2\Omega^*_cL^*}{U^*},\quad
        Pr=\frac{\nu}{\kappa}.
        %\Omega&=\frac{(\Omega_i-\Omega_c)L}{U}, \eta=\frac{Ri}{R_o}.     
\end{equation}
%Therefore, five parameters control the system: $Ra$, $Pr$, $Ro^{-1}$, $\Omega$, and $\eta$. 
Certainly, one can replace several of these five parameters with some other commonly used ones, such as the famous Taylor number $Ta= (1+\eta)^6\Omega^2Ra/16\eta^2(1-\eta)^2Pr$ \citep{zhong_sheared_2023}. In the current study, a practical alternative is the Richardson number measuring the ratio between the buoyancy and shear strength, which reads
\begin{equation}\label{eq: Ri}
     Ri(r)=\frac{N^2}{S^2}=\frac{2r(1-\eta)\partial_r\theta}{(1+\eta)(r\partial_r(u_\varphi/r)+\partial_\varphi u_r/r)^2}.
\end{equation}
Here, $N=\sqrt{{\Omega_c^*}^2r^*\alpha\partial_{r}\theta^*}$ is the buoyancy frequency and $S=r^*\partial_{r}(u^*_\varphi/r^*)+\partial_\varphi u^*_r/r^*$ is the shear stain rate. Note that the definition \eqref{eq: Ri} is a local form. In sheared RBC studies, $Ri$ can be defined directly by the temperature and velocity differences of the two horizontal plates \citep{blass_flow_2020,blass_effect_2021,zhang_twin_2024}. In the current sheared ACRBC system, however, adhering to such a definition is inappropriate due to the non-linear radial distributions of both temperature and velocity base flow. As will be shown later, the local $Ri$ calculated by the base flow changes dramatically along the radial direction, and this non-uniformity is further affected by the radius ratio. Therefore, we will first investigate the properties of local $Ri$ and find a proper global definition afterward.

As reported in ZWS23 with fixed $\eta=0.5$, there exist three regimes in the parameter space $(Ra,\Omega)$: buoyancy-dominated, stable, and shear-dominated regime. In the shear-dominated regime, the shear is much stronger than the buoyancy and the flow behaves like TC flow. Moreover, the solution to the instability problem between the stable regime and the shear-dominated regime can be given by the generalized Rayleigh discriminant \citep{ali_stability_1990,yoshikawa_instability_2013} and has been widely discussed \citep{kang_thermal_2015,meyer_effect_2015,yoshikawa_linear_2015}. Therefore, the effect of the radius ratio on this regime can be reasonably predicted. However, within the buoyancy-dominated regime, the stabilizing influence of shear on buoyancy-driven convection in sheared ACRBC necessitates further investigation into the underlying physics mechanism. Consequently, this paper focuses on the buoyancy-dominated regime, where the flow is quasi-two-dimensional on the $r-\varphi$ plane and becomes gradually stable as the shear increases. Various radius ratios within different $(Ra,\Omega)$ will be considered.

\section{Linear stability analysis}\label{sec:lsa}
Our previous work ZWS23 has revealed that the unstable region of sheared ACRBC is well predicted by the linear theory at $\eta=0.5$. Here we further conduct LSA with respect to different $\eta$, with a particular emphasis on the inhibitory effect of weaker shear on RB instability. As previously mentioned, as $\eta$ approaches 1, the current system tends to wall-sheared RBC. Investigating the similarities and differences in the stability properties of these two scenarios holds significance. 

In the normal LSA approach, the flow field is decomposed into the base flow and perturbation field, i.e.
\begin{equation} \label{eq: psi}
    %\begin{aligned}          
    \psi=\psi_0+\psi'
    %\end{aligned}
\end{equation}
in which $\psi=(\boldsymbol{u},p,\theta)$. The base state solution $\psi_0$ possessed by the equations (\ref{eq: OB}) is stationary and invariant in both axial and azimuthal directions and depends only on $r$, which reads \citep{ali_stability_1990,yoshikawa_instability_2013}:
\begin{equation}\label{eq: baseflow}
    %\begin{aligned}          
    \boldsymbol{u_0}=(Ar+\frac{B}{r},0,0), \quad \theta_0=\frac{ln(r/r_i)}{ln(r_o/r_i)},
    %\end{aligned}
\end{equation}
in which $A=-(1+\eta^2)\Omega/(1-\eta^2), ~B=2r_i^2\Omega/(1-\eta^2)$. Note that $p_0$ can be determined from the other two fields, thus we omit its expression here for simplicity. The perturbation field $\psi'$ is expended into normal modes \citep{meyer_effect_2015,kang_radial_2017}:
\begin{equation}\label{eq: decomp}
    \psi'=\hat{\psi}(r)exp(st+i(n\varphi+kz)),
\end{equation}
in which $\hat{\psi}$ is the radial shape function, $s$ is the temporal growth rate of perturbations, $n$ is the azimuthal mode number and $k$ is the axial wavenumber. Substituting \eqref{eq: psi}-\eqref{eq: decomp} into the governing equations \eqref{eq: OB} and boundary conditions \eqref{eq: BC} and neglecting the high-order terms, one can get eigenfunctions with respect to $\hat{\psi}$. This eigenvalue problem can be numerically solved by discretization on Chebyshev–Gauss–Lobatto collocation points. More details of the LSA approach can be found in ZWS23. In the current work, the number of collocation points ranges from $512$ to $1536$ for good convergence. The LSA is performed over a large Rayleigh number range $10^3\le Ra\le10^9$, a radius ratio range $0.2\le\eta\le0.95$ and a rotating velocity difference range $10^{-3}\le\Omega\le10$. The other two parameters, including the inverse Rossby number and the Prtandl number, are fixed, as $Ro^{-1}=20$ and $Pr=4.3$, according to our previous experiments of ACRBC \citep{jiang_supergravitational_2020, jiang_experimental_2022}.

\begin{figure}
    \centering
    \includegraphics[width=0.5\linewidth]{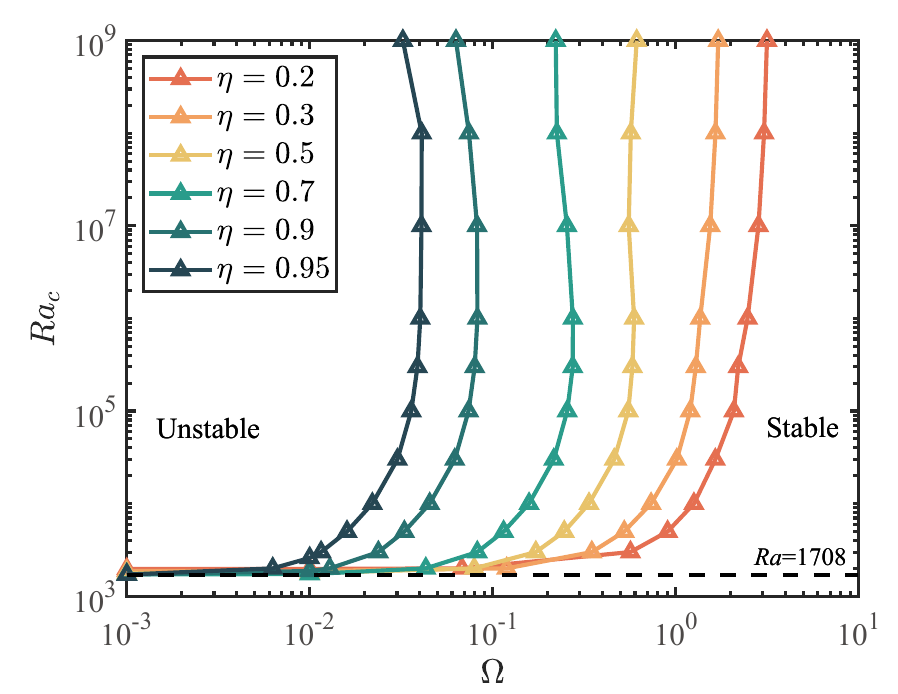}
    \captionsetup{justification=raggedright}
    \caption{The critical Rayleigh number $Ra_c$ versus non-dimensional rotating speed difference $\Omega$ at $\eta=0.2,0.3,0.5,0.7,0.9,0.95$. Each curve means for the marginal states at one radius ratio $\eta$, namely the flow is unstable on the left side of the curve and stable on the right side. The horizontal dashed line represents the critical Rayleigh number $Ra_c=1708$ of RBC.}
	\label{fig: 2}
\end{figure}

Figure \ref{fig: 2} shows the LSA results revealing how the parameter space $(Ra,\Omega)$ is divided into the buoyancy-dominated regime and stable regime at $0.2\le\eta\le0.95$. The variation of the critical Rayleigh number $Ra_c$ with $\Omega$ is consistent with DNS, as will be discussed in section \ref{sec:dns}. When $\Omega\rightarrow0$, there is the onset of unsheared ACRBC, where the critical Rayleigh number $Ra_{c, ACRBC}$ tends to $Ra_{c, RB}=1708$ as $\eta$ gradually approaching $1$ \citep{pitz_onset_2017,wang_effects_2022}. Subsequently, upon introducing shear, the critical Rayleigh number experiences a gradual increment, ultimately leading to an intriguing phenomenon: when $Ra_c\ge10^5$, the marginal-state curve prominently inclines, nearly reaching a vertical orientation. Notably, this trend in the variation of $Ra_c$ with $\Omega$ remains consistent across various radius ratios, while a significant displacement of the marginal-state curve towards the left is observed as $\eta$ progressively escalates. At $Ra=10^7$, the critical $\Omega$ shrinks by almost two orders of magnitude as $\eta$ increases from $0.2$ to $0.95$, which means a much smaller $\Omega$ is needed to stabilize the convection for a larger $\eta$.

\begin{figure}
    \centering
    \includegraphics[width=1.0\linewidth]{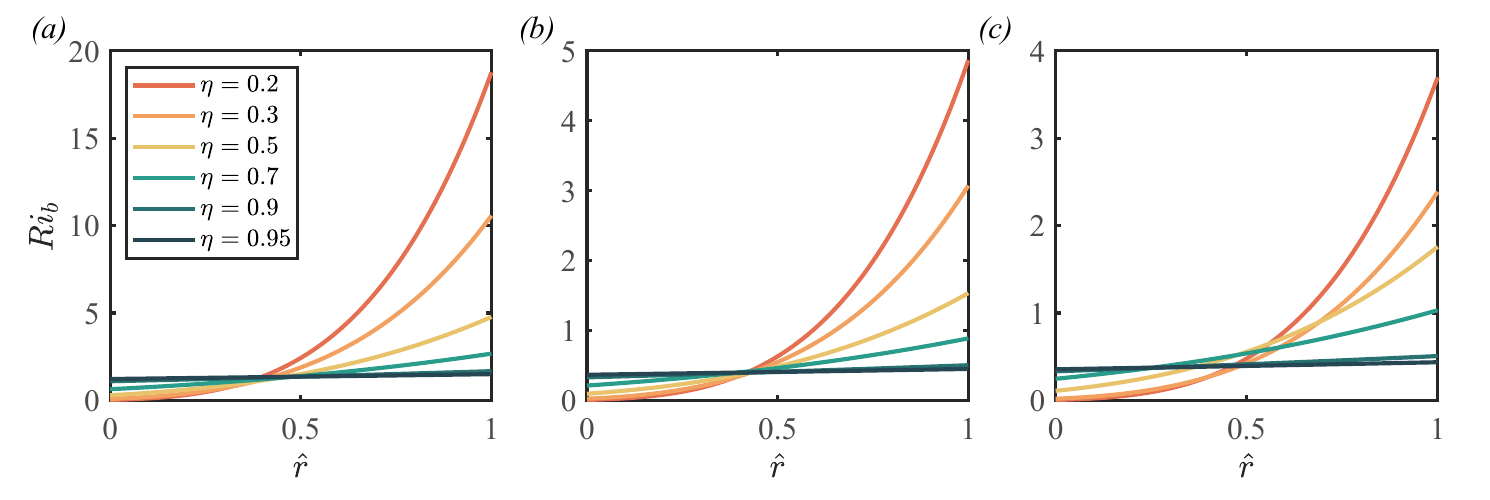}
    \captionsetup{justification=raggedright}
    \caption{The local Richardson number $Ri_b$ defined by the base flow varies with normalized radius $\hat{r}=(r-r_i)$ for the marginal states at different $\eta$ and (\textit{a}) $Ra=10^4$, (\textit{b}) $Ra=10^6$, and (\textit{c}) $Ra=10^7$.}
	\label{fig: 3}
\end{figure}

It is important to note that the smaller $\Omega$ doesn't imply weaker shear when $\eta$ varies. As $r_i=\eta/(1-\eta)$ and $r_o=1/(1-\eta)$, the radii of both inner and outer cylinders increase with $\eta$. Consequently, the velocity differences between two cylinders, i.e. $\Delta_u=\Omega(r_i+r_0)$, may be not small. While it might be natural to substitute $\Delta_u$ for $\Omega$, the results under this parameter do not exhibit consistent behavior. The intrinsic radially non-uniform shear rate distribution in the current system prevents us from simply characterizing global properties using $\Delta_u$. This can be revealed by the local Richardson number calculated by the base flow, namely substituting equation \eqref{eq: baseflow} into \eqref{eq: Ri}, which reads
\begin{equation}\label{eq: Rib}
     Ri_b(\hat{r})=\frac{(1-\eta)^7(1+\eta)}{-8\eta^4ln\eta}\Omega^{-2}(\hat{r}+\frac{\eta}{1-\eta})^4,
\end{equation}
where $\hat{r}=r-r_i\in[0,1]$ is the normalized radius. Obviously, $Ri_b$ increases with $\hat{r}$. For the same $\Omega$, the ratio between the minimum $Ri_b(0)$ at the inner wall and the maximum $Ri_b(1)$ at the outer wall is $\eta^4$. For large $\eta=0.95$, $Ri_b$ is more evenly distributed; while for small $\eta=0.2$, $Ri_b(0)/Ri_b(1)=0.0016$, indicating extremely high inhomogeneity. Note that the radius $r$ cancels in the expression of $N$, thus the buoyancy strength is uniformly distributed and the inhomogeneity of $Ri_b$ mainly comes from the shear. Figure \ref{fig: 3} displays the radial distribution of $Ri_b$ at the marginal state shown in figure \ref{fig: 2}. As $\eta$ increases from $0.2$ to $0.95$, the pronounced non-uniform distribution gradually becomes uniform. A very interesting finding is that the curves representing different radius ratios approximately intersect at one point ($\hat{r}\approx0.45$) for $Ra=10^4$ and $Ra=10^6$; while for $Ra=10^7$, the converging curved lines spread out a little. This implies that the critical $Ri_b$ is almost the same near the middle region for different $\eta$. Therefore, an appropriate global Richardson number can be defined as 
\begin{equation}\label{eq: Rig}
     Ri_g=Ri_b(0.45).
\end{equation}

\begin{figure}
    \centering
    \includegraphics[width=1.0\linewidth]{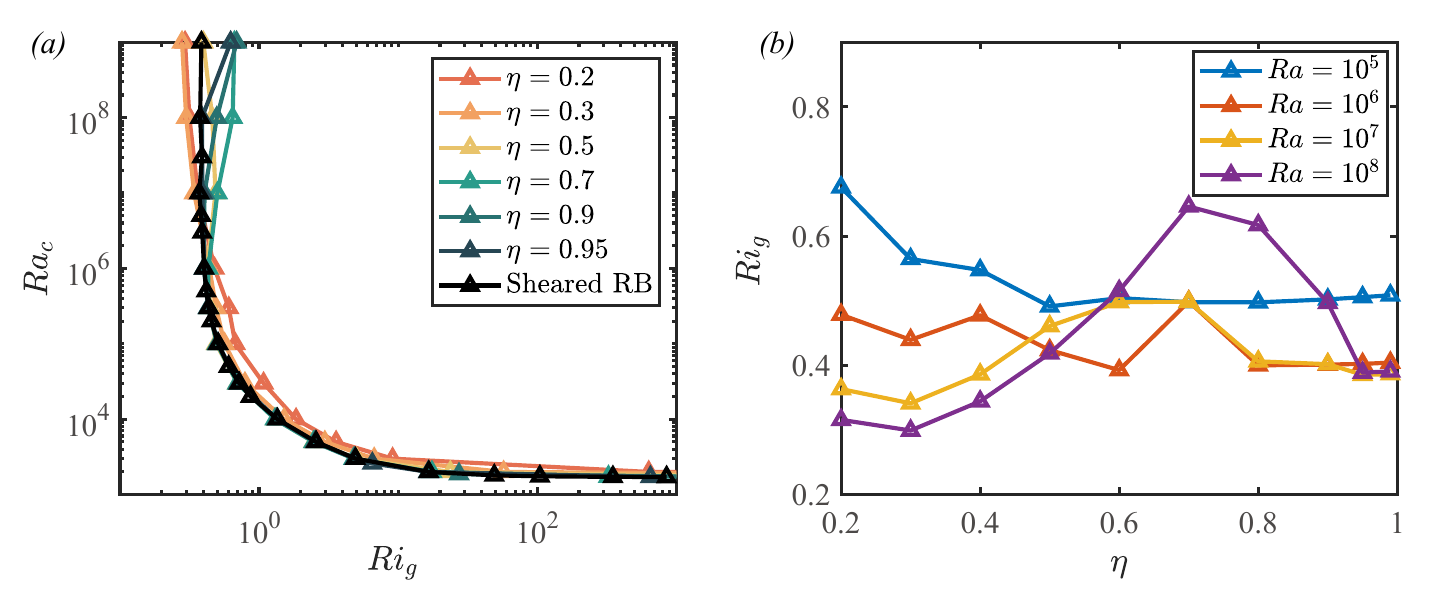}
    \captionsetup{justification=raggedright}
    \caption{(\textit{a}) The critical Rayleigh number $Ra_c$ versus the global Richardson number $Ri_g$ for six radius ratios. The black line means the critical Rayleigh number versus the Richardson number of the transverse rolls in wall-sheared RBC. (\textit{b}) A closer look of (\textit{a}): the critical global Richardson number versus the radius ratio for $Ra=10^5,10^6,10^7,10^8$.  }
	\label{fig: 4}
\end{figure}

With the newly defined $Ri_g$, we convert the marginal-state curves $Ra_c(\Omega)$ to $Ra_c(Ri_g)$, and the results are shown in figure \ref{fig: 4}({\it{a}}). When $Ra_c\leq10^6$, we are delighted to find that all the curves collapse into a single line, except for a small deviation at $\eta=0.2$. When $Ra_c$ exceeds $10^6$, the curves that have collapsed together begin to spread out slightly. We take a closer look in figure \ref{fig: 4}({\it{b}}), picking up four Rayleigh numbers from $10^5$ to $10^8$ to figure out how the critical $Ri_g$ varies with $\eta$. It is shown that for lower $Ra\le10^6$, the critical $Ri_g$ varies little with $\eta$; while for larger $Ra$, the critical $Ri_g$ increases with $\eta$ at first and then decreases. Meanwhile, as $\eta$ approaches $1$, all the curves tend to maintain a positive value rather than zero, which contradicts the absence of a stable state in the three-dimensional wall-sheared RBC \citep{blass_flow_2020,blass_effect_2021}. This inconsistency comes from the fact that the unstable modes of the latter system mainly grow in the spanwise direction, namely the direction perpendicular to the shear and buoyancy, which would be stabilized by strong rotation in sheared ACRBC \citep{jiang_supergravitational_2020}. At large $Ro^{-1}$, the strong Coriolis force suppresses the vertical disturbances, which is a manifestation of the Taylor-Proudman theorem and can also be quantitatively described by the generalized Rayleigh discriminant \citep{bayly_threedimensional_1988, yoshikawa_instability_2013}. In the streamwise direction, we believe that the inhibitory effect of shear on the instability should be similar for both systems. To confirm this statement, we conduct additional LSA on a two-dimensional wall-sheared RBC system and illustrate the results in figure \ref{fig: 4}({\it{a}}) as well. Note that the global Richardson number has a simple definition here, i.e. $Ri_g=g\alpha\Delta^*L^*/{\Delta^*_u}^2$ \citep{blass_flow_2020}. Indeed, the results of wall-sheared RBC agree well with sheared ACRBC, indicating that the streamwise instability mechanisms of the two systems are the same. This also implies that $Ri_g$ defined as \eqref{eq: Rig} serves well as a global control parameter for the current system.

\begin{figure}
    \centering
    \includegraphics[width=1.0\linewidth]{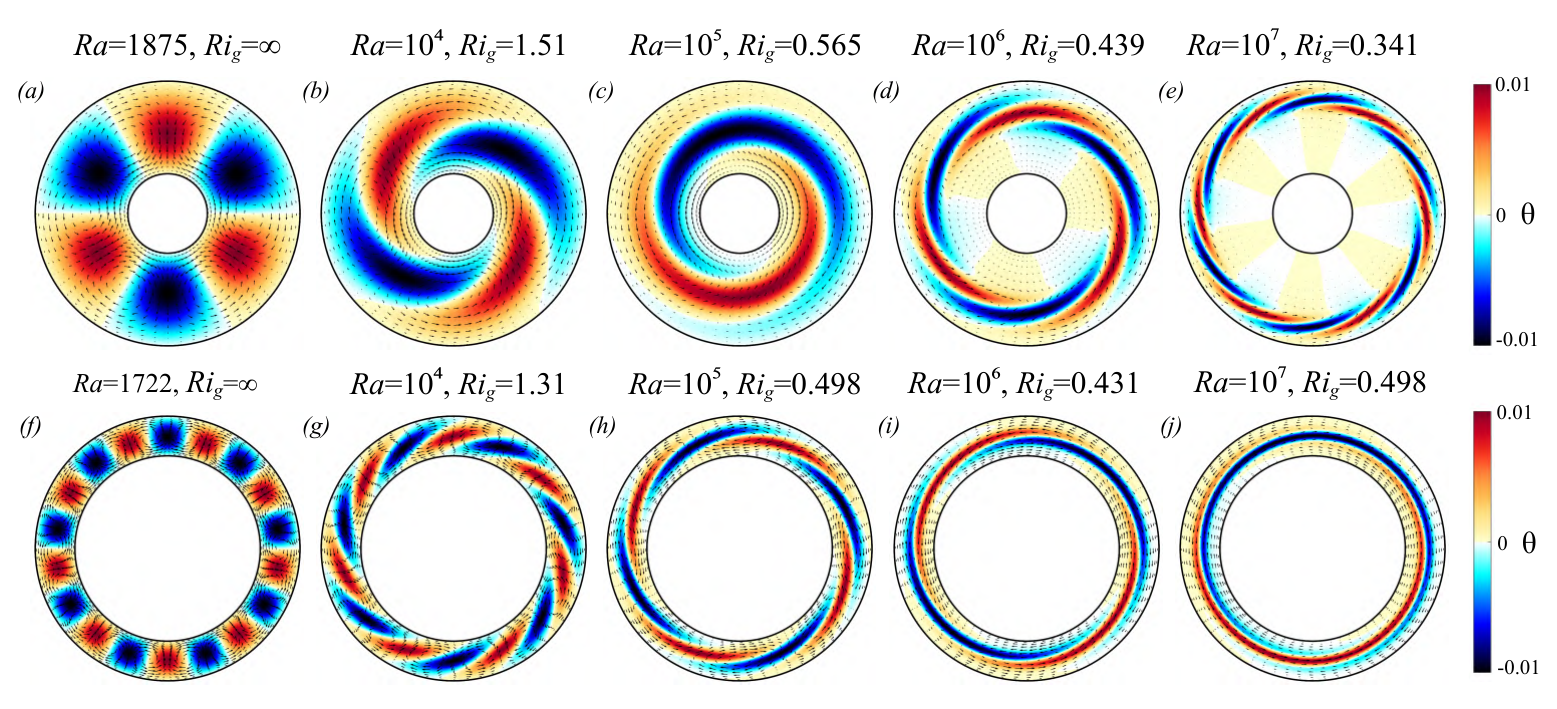}
    \captionsetup{justification=raggedright}
    \caption{Eigenfunctions $(\boldsymbol{u'}, \theta')$ of the critical modes for (\textit{a-e}) $\eta=0.3$ and (\textit{f-j}) $\eta=0.7$ at corresponding Rayleigh numbers and global Richardson numbers.}
	\label{fig: 5}
\end{figure}

\begin{figure}
    \centering
    \includegraphics[width=0.9\linewidth]{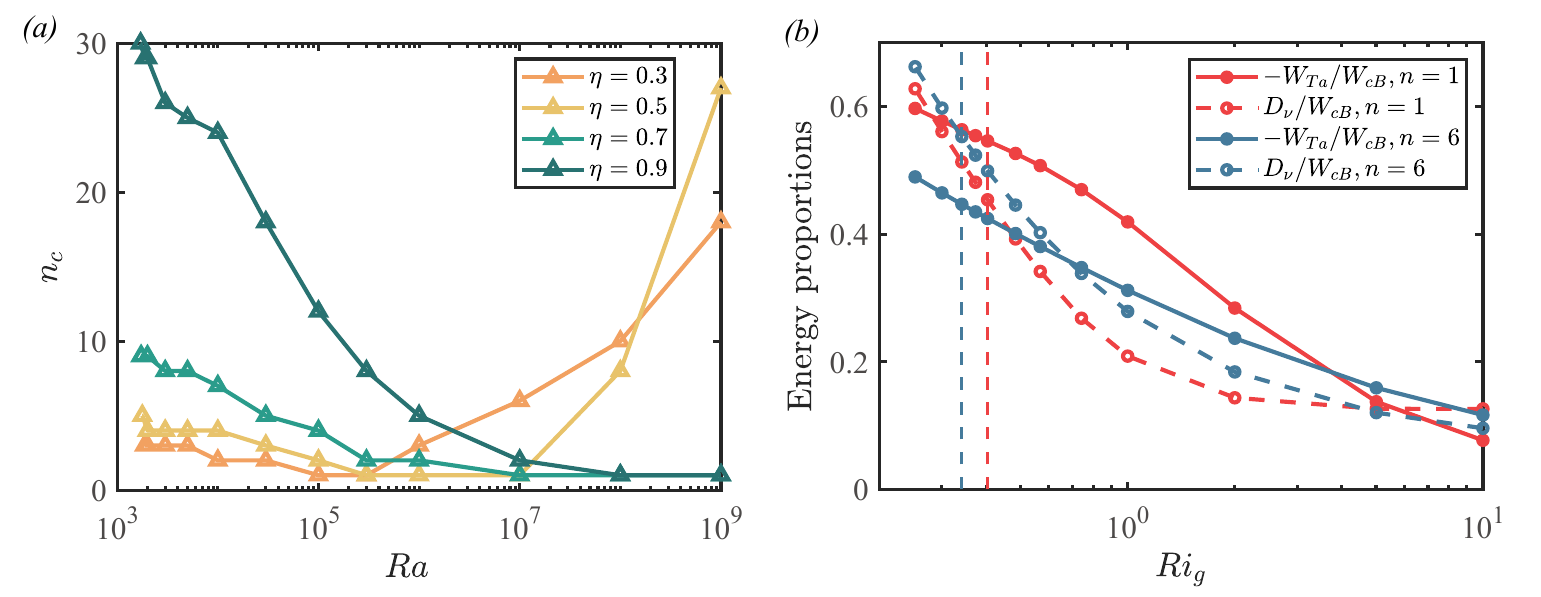}
    \captionsetup{justification=raggedright}
    \caption{(\textit{a}) The critical azimuthal wave number $n_c$ versus $Ra$ at $\eta=0.3,0.5,0.7,0.9$.  (\textit{b}) Variation of energy generation proportions $-W_{Ta}/W_{cB}$ and $D_\nu/W_{cB}$ with the global Richardson number $Ri_g$, for the modes of azimuthal wave number $n=1$ and $n=6$ at $Ra=10^7, \eta=0.3$. The blue vertical dashed line means for the critical $Ri_g$ for $n=6$ and the red vertical dashed line means for the critical $Ri_g$ for $n=1$.} 
	\label{fig: 6}
\end{figure}

Based on the results of wall-sheared RBC, as shown by the black line in figure \ref{fig: 4}({\it{a}}), we can further investigate the deviations at $Ra_c\ge10^7$, namely smaller critical $Ri_g$ appears at around $\eta=0.3$ while larger critical $Ri_g$ appears at around $\eta=0.7$.
In figure \ref{fig: 3}({\it{c}}), the curves do not intersect at a single point at $Ra=10^7$, signifying that the designated value of $\hat{r}=0.45$ may no longer hold its ground as a good representative position of typical instability mode. To investigate the nature of alterations of critical modes at high Rayleigh numbers, the eigenfunctions $(\boldsymbol{u'}, \theta')$ of the critical modes for $\eta=0.3$ and $\eta=0.7$ are displayed in figure \ref{fig: 5}, offering deeper insights into the intricate dynamics at play. When no shear is applied, i.e. $Ri_g=\infty$, there are three hot-cold perturbation roll pairs for small $\eta=0.3$ and nine pairs for large $\eta=0.7$. Such roll pairs will develop into the convection rolls when $Ra>Ra_c$, and the number of roll pairs is determined by the circular roll hypothesis, which implies that the aspect ratio of convection rolls is approximately equal to one \citep{pitz_onset_2017,wang_effects_2022}. As both shear and buoyancy strengths increase along the marginal-state curve, the critical wavenumber gradually decreases for both $\eta=0.3$ and $\eta=0.7$. This is due to the fact that the perturbation modes are elongated in the azimuthal direction under the action of shear, which is similar to the behavior of plumes under shear \citep{goluskin_convectively_2014,blass_flow_2020}. The perturbation roll pairs are slightly off-center towards the inner wall, corresponding to the chosen radius $\hat{r}=0.45$ for the global Richardson number. Till $Ra=10^5$, there is only one roll pair in the case of $\eta=0.3$. An interesting phenomenon is discovered as $Ra$ increases to $10^6$: the critical mode moves towards the outer wall and the wavenumber begins to increase with $Ra$. However, for $\eta=0.7$, this phenomenon does not happen. The roll pairs are still located near the middle and the wavenumber remains unity when $Ra\ge10^6$. In figure \ref{fig: 6}({\it{a}}), we summarized the variation of critical azimuthal wavenumber $n_c$. It is observed that for smaller $\eta$, $n_c$ begins to increase with $Ra$ earlier after decreasing to unity. Within the considered range of $Ra$, such re-increase of wavenumber is absent for large $\eta=0.7$ and $0.9$, but it may occur at much higher $Ra$.

The physical interpretation for the above phenomena is two-fold. Firstly, the current annular system inherently constrains the infinite growth of azimuthal wavelength, which does not exist in wall-sheared RBC. Consequently, when $n_c$ decreases to 1 and $Ra$ further increases, the critical shear strength, originally applicable to the modes with longer wavelength, no longer applies to the mode of which the wavenumber remains unity. The elongation of the perturbation filed for this mode does not further increase, resulting in a smaller corresponding critical shear strength. This explains the phenomenon of larger $Ri_g$ at around $\eta=0.7$ and higher $Ra$, as depicted in figure \ref{fig: 4}({\it{b}}). Meanwhile, in figure \ref{fig: 4}({\it{a}}), this can also explain the fact that the curves of large radius ratios deviate sequentially to larger $Ri_g$ from the marginal-state curve of wall-sheared RBC when $Ra\ge10^7$. Secondly, the radially non-uniform distribution of shear strength in the current system causes the most unstable mode to shift toward the outer wall. As seen in figure \ref{fig: 3}, for small $\eta$, the shear strength near the outer wall is significantly smaller than that from the center to the inner wall. Considering the stabilizing effect of shear on unstable modes, when $Ra$ is sufficiently large (corresponding to a longer distance between the two walls), the unstable modes tend to develop preferentially near the outer wall. At this point, the critical shear strength at $\hat{r}=0.45$ overestimates the dominated mode near $\hat{r}=1$. This elucidates the phenomenon of smaller $Ri_g$ at around $\eta=0.3$ and higher $Ra$, as observed in figure \ref{fig: 4}.

The above discussion can be further demonstrated from the perspective of energy. The kinetic energy equation of perturbations is expressed as \citep{yoshikawa_instability_2013,yoshikawa_linear_2015,meyer_effect_2015}:
\begin{equation}\label{EnergyEqu}
    \frac{dK}{dt}=W_{Ta}+W_{cB}-D_\nu,
\end{equation}
where $K$ is the kinetic energy, $W_{Ta}$ is the rate of energy exchanged from the inertial shear flow, $W_{cB}$ is the power of centrifugal buoyancy and $D_\nu$ is the energy dissipation rate due to viscosity, respectively. Detailed expressions for each of the above terms can be found in equation (3.2) of our previous paper ZWS23. Note that $W_{Ta}$ is usually negative in the ACRBC system, implying that the energy released by centrifugal buoyancy is consumed by both dissipation and azimuthal shear flow. We select the cases at $\eta=0.3$ and $Ra=10^7$, concentrating on how the energy generation terms of the two kinds of modes with azimuthal wave number $n=1$ (located in the middle) and $n=6$ (located closer to the outer cylinder with stronger shear) vary with increasing shear, and the results are illustrated in figure \ref{fig: 6}({\it{b}}). Here we consider the proportions of energy generation terms relative to the buoyancy term, i.e. $-W_{Ta}/W_{cB}$ and $D_\nu/W_{cB}$, the sum of which reaching one means for the marginal state. As shown in figure \ref{fig: 6}({\it{b}}), both the inertial term and viscous term consume more proportions of the energy of buoyancy for $n=1$ and $n=6$ with the shear enhancement, indicating that the shear suppresses the growth of instability induced by buoyancy. When comparing the modes with $n=1$ and $n=6$, we discover that under weak shear (high $Ri_g$), the proportions of total energy consumption are close between the two modes. As $Ri_g$ tends to critical value for $n=1$, as denoted by the red vertical dashed line in figure \ref{fig: 6}({\it{b}}), the viscous proportion of the mode with $n=6$ is a bit larger than that of the mode with $n=1$, but the inertial proportion is much smaller for the former, making the corresponding mode unstable. That is, the outward shifting of the perturbation mode is advantageous for reducing the energy converting to the shear flow, thus in turn promoting the development of the mode. Therefore, the critical mode changes from the middle mode to the outward mode with smaller critical $Ri_g$, as denoted by the vertical blue dashed line in figure \ref{fig: 6}({\it{b}}), which is consistent with our previous reasoning.

\section{Direct numerical simulation}\label{sec:dns}
\begin{figure}
    \centering
    \includegraphics[width=0.9\linewidth]{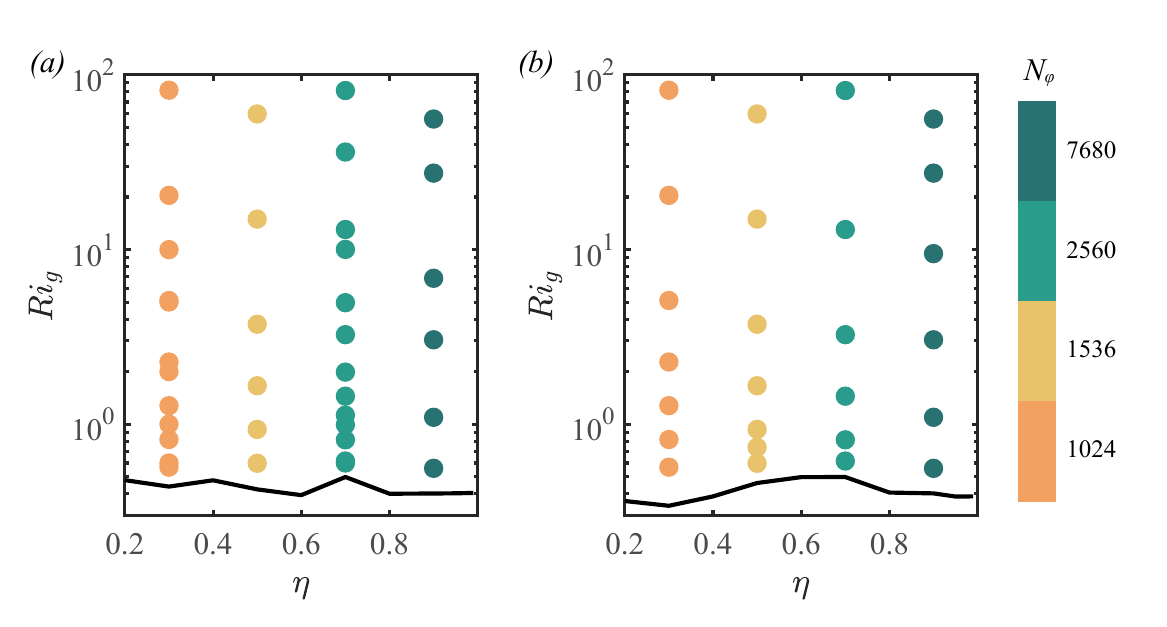}
    \captionsetup{justification=raggedright}
    \caption{ The distribution of main simulation parameters and the corresponding azimuthal resolutions $N_\varphi$ in the $(\eta, Ri_g)$ domain under (\textit{a}) $Ra=10^6$ and (\textit{b}) $Ra=10^7$. The black solid lines denote the marginal state.}
	\label{fig: simupara}
\end{figure}

Based on the LSA results, fully nonlinear numerical simulations are performed using an energy-conserving second-order finite-difference code AFiD \citep{ van_der_poel_pencil_2015, zhu_afid-gpu_2018}, which has been validated many times in the literature \citep{verzicco_finite-difference_1996, Ostilla_2014_Exploring, jiang_supergravitational_2020,jiang_experimental_2022}. As in the buoyancy-dominated regime, the flow in the sheared ACRBC is quasi-two-dimensional \citep{jiang_supergravitational_2020, zhong_sheared_2023}, the simulations are performed on a two-dimensional cyclic cross-section, with the radius ratio $\eta\in[0.3,0.9]$. Two Rayleigh numbers $Ra=10^6$ and $10^7$ are selected and the global Richardson number $Ri_g$ varies from the critical value to $10^2$, as shown in figure \ref{fig: simupara}. The critical $Ri_g$ predicted by LSA has been validated by additional cases in the stable regime, which are not presented in the figure for simplicity. We have performed the posterior check on the relevant scales including the Kolmogorov scale and the Batchelor scale to guarantee adequate resolutions \citep{silano_numerical_2010}. Meanwhile, the Courant–Friedrichs–Lewy (CFL) conditions are used as $CFL\le0.7$ to ensure computational stability \citep{ostilla_optimal_2013, van_der_poel_pencil_2015}. Moreover, enough simulation time is ensured to limit the error in the statistics. All the numerical details of the unstable cases are illustrated in the Appendix.

\subsection{Initial development}
\begin{figure}
    \centering
    \includegraphics[width=0.8\linewidth]{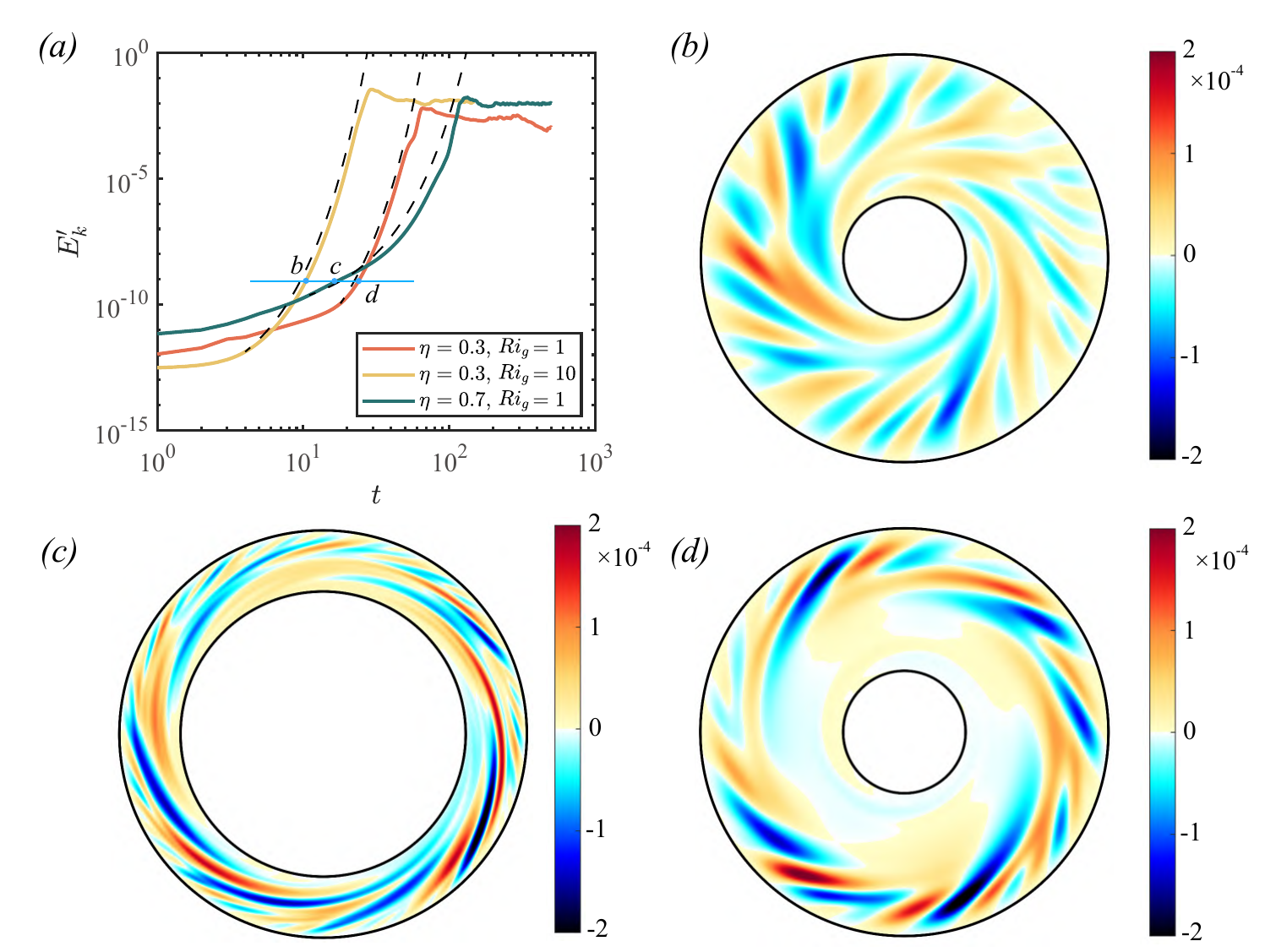}
    \captionsetup{justification=raggedright}
    \caption{(\textit{a}) Time series of the mean perturbation energy $E'_k=\langle|\boldsymbol{u'}|^2\rangle_{V}/2$ for three cases with $(\eta, Ri_g)=(0.3, 1)$, $(0.3, 10)$ and $(0.7, 1)$ at $Ra=10^7$. The dashed lines represent the predictions of LSA. (\textit{b,c,d}) The perturbation temperature fields at the instants marked in (\textit{a}) for corresponding cases.}
	\label{fig: init}
\end{figure}

In the DNS, small random perturbations are added to trigger the flow development. When the Rayleigh number $Ra$ is larger than the critical $Ra$ (or the rotating angular speed difference $\Omega$ is smaller than the critical $\Omega$), the perturbations will grow up linearly at first. To investigate the initial development, we calculate the perturbation energy $E'_k=\langle|\boldsymbol{u'}|^2\rangle_{V}/2$ from the instantaneous velocity fields and depict its time evolution for three typical cases, i.e. $(\eta, Ri_g)=(0.3, 1)$, $(0.3, 10)$ and $(0.7, 1)$, in figure \ref{fig: init}({\it{a}}). Meanwhile, we draw the LSA results calculated by the growth rate of the linear fastest-growing mode for each case, as indicated by the dashed lines. It can be seen that after the mode with the highest growth rate dominates, the perturbation energy grows in line with the predictions given by LSA until it approaches the peak, where the linear mode saturates and the non-linear effects begin to make sense. Therefore, the instability and initial development of the flow field for different radius ratios in ACRBC can be well described by the linear theory.

Moreover, we have performed checks on the outward displacement of critical modes given by LSA. Figures \ref{fig: init}({\it{b-d}}) show the instantaneous temperature perturbation fields that are denoted in figure \ref{fig: init}({\it{a}}). Different initial modes can be found in the linear stage. For $\eta=0.3$, when the shear is weak ($Ri_g=10$), the perturbations develop in the entire space. Since this is not a critical mode, many pairs of hot and cold plumes can be observed. These plumes are elongated in the azimuthal direction by shear, which is similar to the modes obtained by LSA. Under the strong shear ($Ri_g=1$), however, perturbations develop only in parts close to the outer cylinder, while perturbations close to the inner cylinder are suppressed. Correspondingly, in a large radius ratio system under the same strong shear $(\eta=0.7, Ri_g=1)$, the perturbations still occupy the whole domain. These phenomena are consistent with the LSA results.

\subsection{Flow structures}\label{sec: flow}
\begin{figure}
    \centering
    \includegraphics[width=1.0\linewidth]{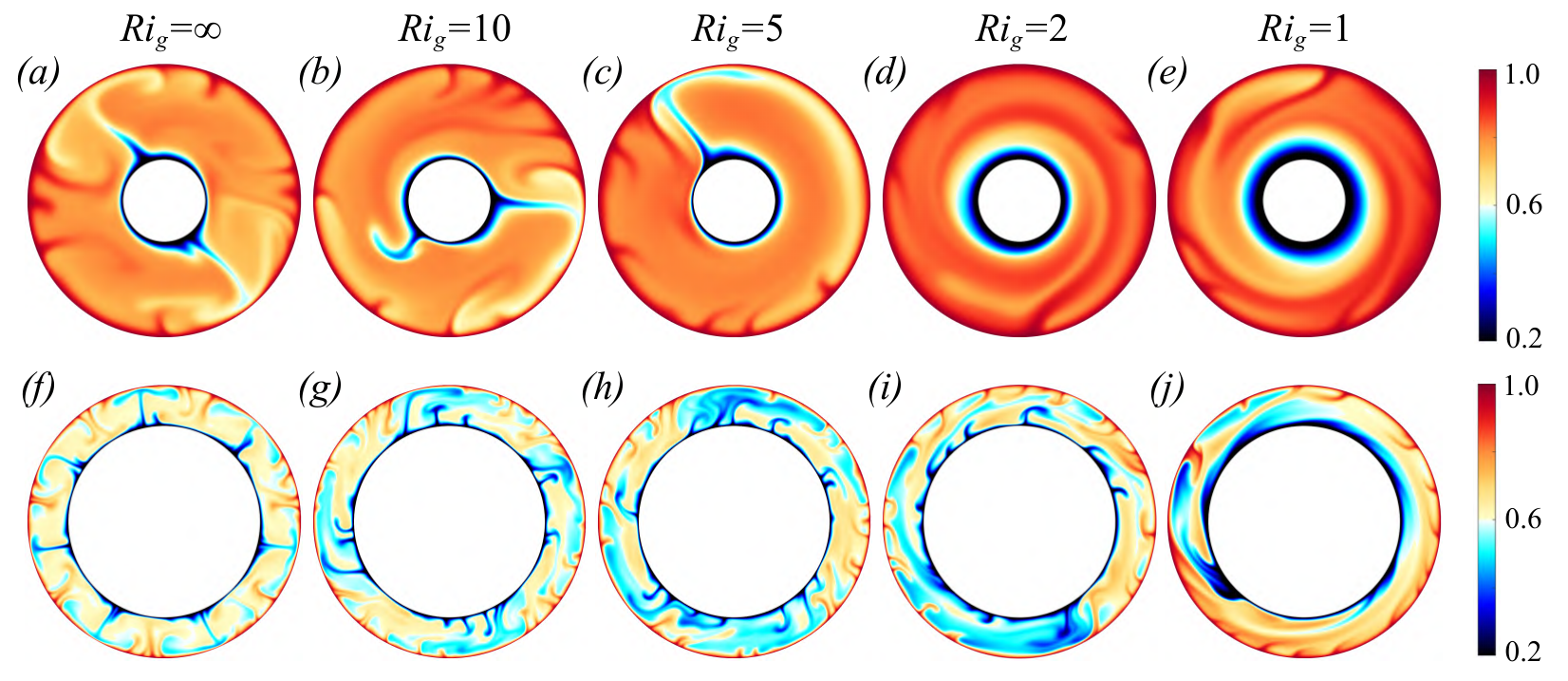}
    \captionsetup{justification=raggedright}
    \caption{Typical snapshots of the instantaneous temperature field on the $r\varphi$ plane at $Ri_g=\infty,10,5,2,1$ under (\textit{a-e}) $ \eta=0.3$ and (\textit{f-j}) $\eta=0.7$. $Ra=10^6$.}
	\label{fig: 8}
\end{figure}

When the perturbations develop further to form convection, a statistically steady state can be found. In this section, we focus on the flow structures in this state. Figure \ref{fig: 8} shows some typical snapshots of the instantaneous temperature field on the $r\varphi$ plane with increasing shear strength under $\eta=0.3$ and $0.7$ at $Ra=10^6$. Without shear, two pairs of convection rolls appear at $\eta=0.3$ while seven pairs appear at $\eta=0.7$. The fact that more pairs of convection rolls form at larger $\eta$ has been confirmed by previous LSA. Due to the Coriolis force, the cold and hot plumes turn to the right when crossing the bulk region, breaking the symmetry of one roll pair. The single roll of a pair in the plume deflection direction becomes larger and the other becomes smaller \citep{wang_effects_2022}. When the shear is applied, the movement direction of the two walls aligns precisely with the rotation direction of the larger roll, thereby further enhancing the asymmetry. Consequently, as the shear strengthens, the convection rolls gradually diminish until they cease to exist.

%Therefore, due to the asymmetry of the two rolls, the bigger roll takes more area and has a larger impact on the boundaries than the other roll, leading to a negative $Nu_\omega$ at a small shear strength.
Since there are fewer convection rolls for a small radius ratio, they quickly disappear when shear becomes stronger. For $\eta=0.3$, only one strong cold plume and several hot plumes remain at $Ri_g=5$, as shown in figure \ref{fig: 8}({\it{c}}). The number of hot plumes is greater than that of cold plumes because the surface of the outer cylinder is much larger than the surface of the inner cylinder, which is one of the manifestations of the asymmetry in ACRBC. With the further enhancement of the shear, the cold plume disappears, while significant long tilting hot plumes derive from the outer cylinder. This phenomenon again validates the outward shift of the critical modes discovered in the LSA, which indicates that the thermal convection pattern is also affected by the inhomogeneous distribution of the shear, and the influence is more pronounced at small radius ratios.

Under $\eta=0.7$, since more convection roll pairs exist without shear, their disappearance occurs at smaller $Ri_g$. Until $Ri_g=1$, although no significant convection rolls are present, there are still many plumes detached from both the inner and outer cylinders, as shown in figure \ref{fig: 8}({\it{j}}). This is partly due to the large inner wall area of the system with large $\eta$, which therefore allows for more plumes to be generated, and partly because the shear effect is more uniform, which means that the shear on the inner cylinder side is not as strong as that in the case with small $\eta$. When the shear is further enhanced, the plumes on the inner and outer cylinder surfaces are further suppressed as well.

%From a heat transfer perspective,  the number of plumes decreases more slowly and the plume disappears later at large radius ratios, which is the reason why heat transfer is suppressed later.

\begin{figure}
    \centering
    \includegraphics[width=0.9\linewidth]{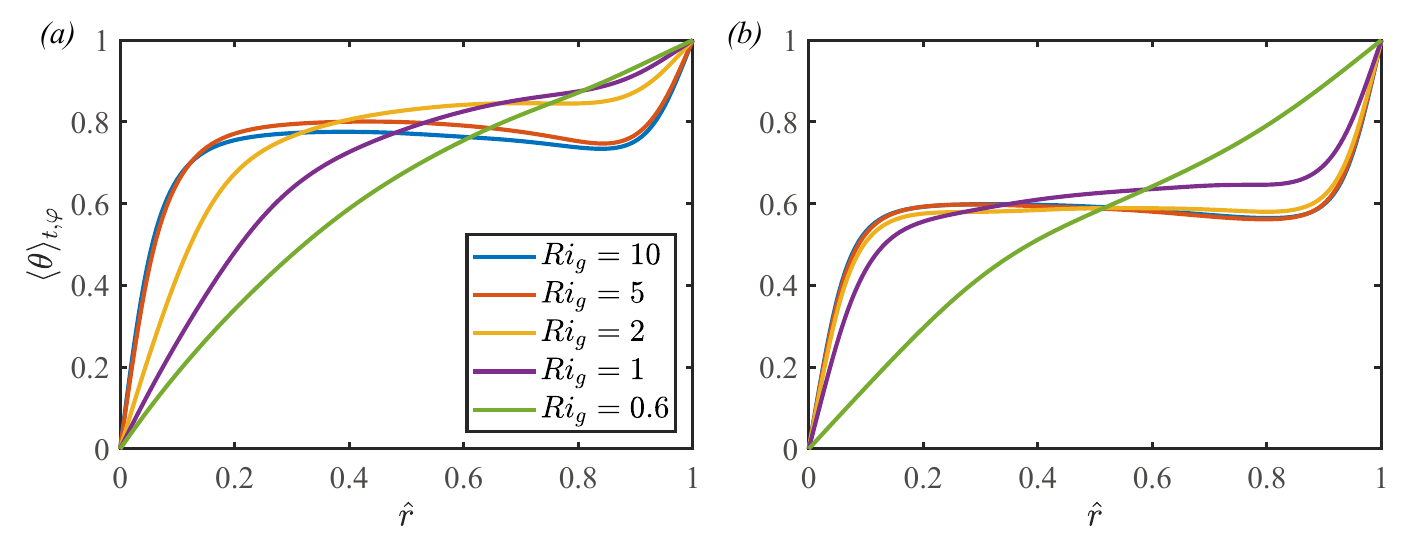}
    \captionsetup{justification=raggedright}
    \caption{Radial distribution of azimuthally and time-averaged temperature $\langle\theta\rangle_{t,\varphi}$ at different shear strength for (\textit{a}) $\eta=0.3$ (\textit{b}) $\eta=0.7$. $Ra=10^6$.}
	\label{fig: 9}
\end{figure}

In the snapshots of the temperature field, differences in the bulk temperatures for different $\eta$ are another concern. For ACRBC without shear, the bulk temperature increases from $\theta_m=0.5$ as $\eta$ decreases from $1$. The enhancement of bulk temperature is caused by the asymmetry of ACRBC in the radial direction, and the effect of radius ratio on the asymmetric temperature distribution is well described by \cite{wang_effects_2022}. In the sheared ACRBC system, this asymmetry has more profound implications for the flow dynamics. In figure \ref{fig: 9} we plot the averaged temperature profiles of different $Ri_g$ under $\eta=0.3$ and $0.7$. It can be seen that at weak shear, the bulk temperature at $\eta=0.3$ is larger than that at $\eta=0.7$. With the increase of shear strength, the uniform bulk temperature gradually increases, meanwhile, the uniform bulk area shifts towards $\hat{r}=1$. For the small $\eta=0.3$,  a significant increase of bulk temperature and the corresponding shifting happen at a larger $Ri_g=2$, where the cold plumes totally disappear, as shown in figure \ref{fig: 8}({\it{d}}). While for $\eta=0.7$, the bulk temperature remains nearly constant until $Ri_g=1$, indicating the robust bulk convective mixing. Afterward, the flow suddenly evolves to the laminar and non-vortical state. Again, this is consistent with the LSA results, illustrating that the inhomogeneity of the shear distribution affects the sheared ACRBC at different radius ratios with different intensities in various aspects including stability and flow structures.

\subsection{Global transportation}
The different flow structures for different $\eta$ further affect the global transportation in sheared ACRBC. The heat transfer efficiency and the momentum transfer efficiency in the statistically steady state are measured by two Nusselt numbers: $Nu_h$ and $Nu_\omega$, defined as the ratios of the corresponding fluxes of the current system to the fluxes in the laminar and nonvortical flow case \citep{eckhardt_torque_2007,wang_effects_2022, zhong_sheared_2023}:
\begin{equation}
	\begin{aligned}
		Nu_h&=\frac{\sqrt{RaPr}\langle u_r\theta\rangle_{t,\varphi,z}-\partial_r\langle\theta\rangle_{t,\varphi,z}}{(rln(\eta))^{-1}},\\
		Nu_\omega&=\frac{r^3[Ra/Pr\langle u_r\omega\rangle_{t,\varphi,z}-\sqrt{Ra/Pr}\partial_r\langle\omega\rangle_{t,\varphi,z}]}{2B},\\
	\end{aligned}
\end{equation} 
where $\omega=u_\varphi/r$ is the angular velocity of the fluid, and $B$ is the parameter of the base flow defined in equations (\ref{eq: baseflow}). $\langle \cdot \rangle_{t,\varphi,z}$ represents the temporal-, azimuthal- and axial-averaged value. In ACRBC without shear, i.e. $\Omega=0$ or $Ri_g=\infty$, $Nu_h$ decreases with decreasing $\eta$ for a fixed $Ra$ \citep{wang_effects_2022}. Meanwhile, it is known that shear will suppress the heat transfer efficiency as well \citep{blass_flow_2020,zhong_sheared_2023}. When shear is introduced in ACRBC, what would be the difference in the relationship of $Nu$ with shear strength at different $\eta$? To make a reasonable comparison of shear strengths at systems with different $\eta$, we still adopt the global Richardson number $Ri_g$ to represent the shear strength here. 

\begin{figure}
    \centering
    \includegraphics[width=1.0\linewidth]{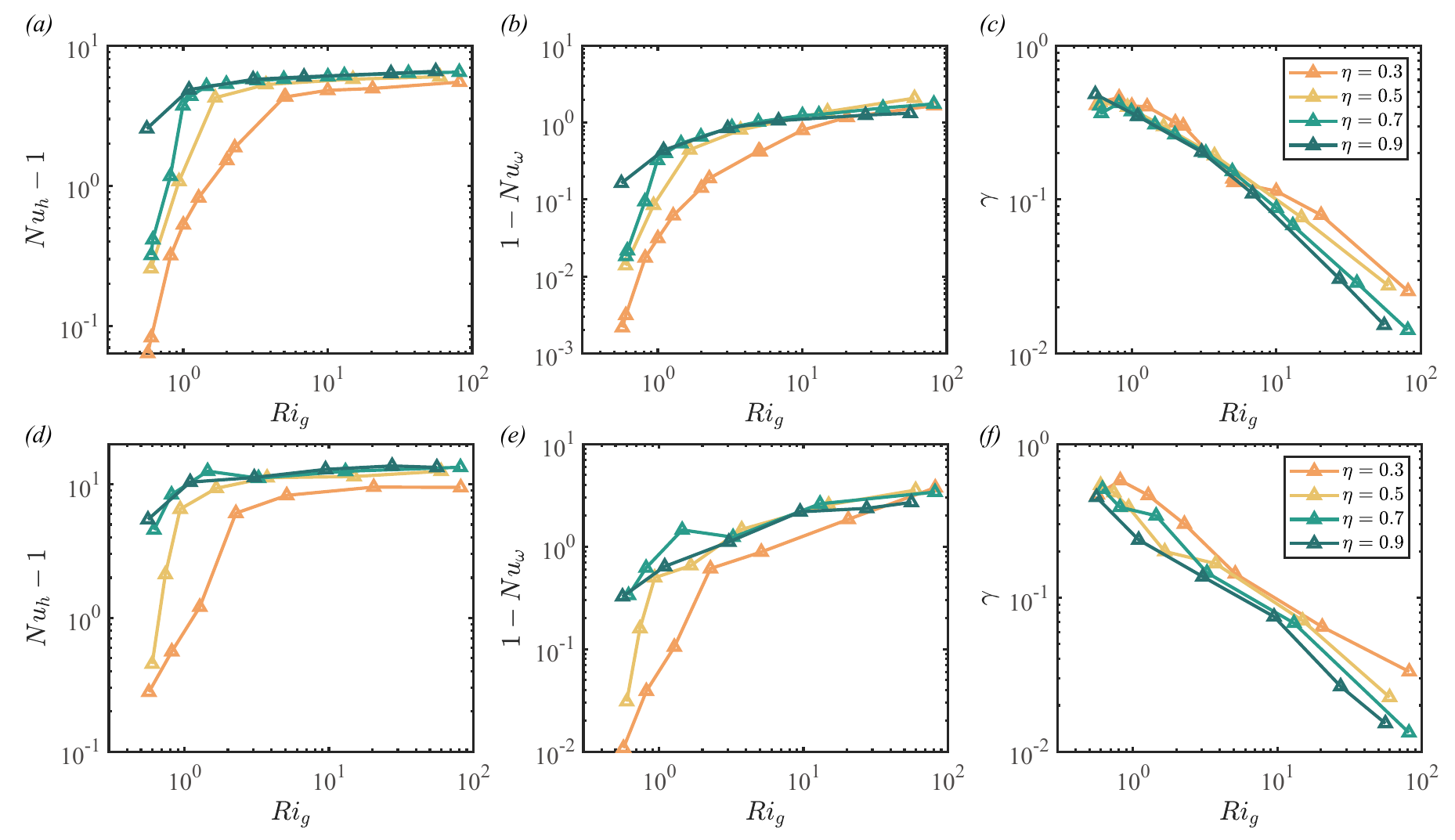}
    \captionsetup{justification=raggedright}
    \caption{Variation of (\textit{a,d}) $Nu_h$ and (\textit{b,e}) $Nu_\omega$ (\textit{c,f}) $\gamma$ with $Ri_g$ at $\eta=0.3$, $0.5$, $0.7$ and $0.9$. The data in the first row (\textit{a-c}) are calculated at $Ra=10^6$, while the data in the second row (\textit{d-f}) are calculated at $Ra=10^7$.}
	\label{fig: 7}
\end{figure}
The variations of the two Nusselt numbers with $Ri_g$ at different $\eta$ are illustrated in figures \ref{fig: 7}({\it{a,b}}) for $Ra=10^6$ and in figures \ref{fig: 7}({\it{d,e}}) for $Ra=10^7$. With the increase of shear strength (decreasing $Ri_g$), $Nu_h$ decreases slowly at first and then rapidly when the flow approaches the marginal state. The value $Nu_h-1$ in the figures reflects the extent of heat transfer enhancement compared to heat conduction. This trend holds for different radius ratios and the two Rayleigh numbers. The transition in the rate of decline of $Nu_h$ can be clearly seen in the logarithmic coordinate system of figures \ref{fig: 7}({\it{a,d}}), which exactly corresponds to the vanishing of convection rolls, as shown in figure \ref{fig: 8}. For example, the rapid decrease of $Nu_h$ occurs when $Ri_g<5$ for $\eta=0.3$ and $Ri_g<1$ for $\eta=0.7$. Therefore, the shear has smaller effects on the heat transfer before the break of large convection rolls. This means that for large $\eta$ with robust convection, a nearly constant $Nu_h$ can hold for a large range of $Ri_g$, as can be seen in figures \ref{fig: 7}({\it{a,d}}).

%Based on the previous analysis of linear stability, fewer perturbation roll pairs appear in cases with smaller $\eta$ at $Ri_g=\infty$, so as shear is enhanced, the number of perturbation roll pairs decreases earlier to one. This is consistent with the trend of heat transfer with shear. We will continue to discuss this point in the subsequent section \ref{sec: flow} on flow structures.

In the buoyancy-dominated regime of sheared ACRBC, $Nu_\omega$ is smaller than $1$, which means that the drag on the boundaries is smaller than the drag of base flow \citep{zhong_sheared_2023}. For a weak shear, $Nu_\omega$ even becomes negative, indicating that the large convection rolls push the two cylinders to rotate. Therefore, in the figures \ref{fig: 7}({\it{b,e}}), $1-Nu_\omega$ is considered, which represents the role of thermal convection on wall motion. When shear is weak, the values of $Nu_\omega$ are close for different $\eta$. With enhanced shear, $1-Nu_\omega$ yields the same trend as $Nu_h-1$, namely decreasing slowly at first and rapidly afterward. The transition similarly occurs when the convection rolls disappear. Therefore, the global convection mode holds great significance for both heat and momentum transfer in sheared ACRBC. 

To further investigate the relationship between heat and momentum transfer, we again adopt the perspective of energy. In the dimensional form, as $u^*_\varphi/r^*\ll\Omega^*_c$ in the buoyancy-dominated regime, the global energy balance of our system can be derived from the equation (\ref{eq: OB}) \citep{eckhardt_torque_2007, wang_effects_2022, zhong_sheared_2023}:
\begin{equation}\label{NuEqu}
    \varepsilon-\varepsilon_{lam}=\nu^{3}L^{*-4}[\sigma_r^{-2}Ta(Nu_\omega-1)+f(\eta)Pr^{-2}Ra(Nu_h-1)],
\end{equation}
where $\varepsilon=\nu\langle(\partial_iu^*_j+\partial_ju^*_i)^2\rangle_{V,t}$ is the mean energy dissipation rate, $\varepsilon_{lam}$ is the mean energy dissipation rate of the laminar and nonvortical flow, $\sigma_r=(1+\eta)^4/16\eta^2$ is the quasi-Prandtl number, and $f(\eta)=\frac{2(\eta-1)}{(1+\eta)ln(\eta)}$ is a correction factor. The two terms on the right side represent the energy injected by shear and buoyancy, respectively. As the momentum Nusselt number $Nu_\omega<1$, the first term on the right side is negative, indicating that the shear consumes energy and only the buoyancy provides. The ratio $\gamma$ of the energy consumed by shear and the energy injected by buoyancy reads
\begin{equation}
	\gamma=\frac{\sigma_r^{-2}Ta(1-Nu_\omega)}{f(\eta)Pr^{-2}Ra(Nu_h-1)}=\frac{-8\eta^2ln\eta(1-Nu_\omega)}{(1+\eta)(1-\eta)^3(Nu_h-1)}Pr\Omega^{2}.
\end{equation}
Figures \ref{fig: 7}({\it{c,f}}) show that how $\gamma$ varies with $Ri_g$ at different radius ratios in $Ra=10^6$ and $10^7$, respectively. Basically, $\gamma$ increases with decreasing $Ri_g$ in an approximate power law relation. Interestingly, for $Ra=10^6$ and small $Ri_g$, the curves representing different radius ratios, which are separated in the other two figures, collapse together in the $\gamma-Ri_g$ relation, indicating that the energy allocation rules in the flow closed to the stable state are similar for different $\eta$. Meanwhile, this implies that $Ri_g$ is not only applied for the initial linear instability but also the fully developed flow field. At larger $Ri_g$, the shear is weak and the convection rolls are strong, $\gamma$ is larger for smaller $\eta$. The reason for this may be that the nonlinearities of large convection rolls introduce new factors related to $\eta$ to come into play, such as the curvature, Coriolis force, and the zonal flow \citep{wang_effects_2022}. Consequently, single $Ri_g$ cannot completely describe the effect of different radius ratios on the heat and momentum transfer of the system. For the cases with $Ra=10^7$, in which the convection rolls are more intense, the curves representing different $\eta$ are always slightly separated. As discussed in section \ref{sec:lsa}, for high $Ra$, the outward shift of critical mode and the confinement of azimuthal wavelength make the critical $Ri_g$ vary slightly. Nevertheless, considering the comprehensive results above, $Ri_g$ behaves well to characterize the overall trend of heat and momentum transfer of sheared ACRBC at different radius ratios.

%\begin{figure}
%    \centering
%    \includegraphics[width=1.0\linewidth]{fig_eps/fig7_new.pdf}
%    \captionsetup{justification=raggedright}
%    \caption{Variation of (\textit{a}) $Nu_h$ and (\textit{b}) $Nu_\omega$ (\textit{c}) $\gamma$ with $Ri_g$ at $\eta=0.3$, $0.5$, $0.7$ and $0.9$. $Ra=10^6$.}
%	\label{fig: 7}
%\end{figure}
\section{Conclusion}\label{sec:con}
In the present study, we investigate the effect of radius ratio on the sheared ACRBC system
by linear stability analysis and direct numerical simulations. Guided by the description of \cite{zhong_sheared_2023}, since the temperature only works as a passive scalar in the shear-dominated regime, we concentrate on the buoyancy-dominated regime of sheared ACRBC, where the quasi-two-dimensional thermal convection is gradually suppressed by increasing imposed shear. Through the linear stability analysis, we observe that as the radius ratio $\eta$ increases from $0.2$ to $0.95$, the marginal-state curve $Ra_c(\Omega)$ shifts along the $-\Omega$ direction, which means a smaller $\Omega$ is required to stabilize the flow. Considering the inhomogeneity of the shear strength distribution due to the geometric asymmetry, a global Richardson number $Ri_g$ is defined in terms of the most representative local Richardson number. With the newly defined $Ri_g$, the marginal-state curves under different radius ratios are collapsed together in the parameter domain $(Ra, Ri_g)$, also consistent with the marginal-state curve $Ra_c(Ri)$ of the wall-sheared RBC in the streamwise direction. This demonstrates that the stabilization mechanism in the direction of shear flow is identical for the two systems. In addition, due to the geometrical limitation of the maximum azimuthal wavelength, the marginal-state curves in sheared ACRBC are offset from that of the wall-sheared RBC under high-intensity shear. 

The equivalent aspect ratio of the system at low radius ratios is smaller, which allows the system to accommodate fewer roll pairs according to the circular roll hypothesis \citep{wang_effects_2022}. When shear is applied, this causes the convection rolls as well as the thermal plumes in the system to disappear more quickly, thus allowing the heat transfer to be drastically suppressed in advance. Meanwhile, the strong asymmetry of the small radius ratio system causes significant disparity in the quantities of hot and cold plumes along with temperature elevation in the bulk region, and the imposition of shear further exacerbates these effects. Interestingly, even if flow structures differ, the percentage of buoyant energy consumed by shear varies consistently with $Ri_g$ for systems with different radius ratios. This, in turn, indicates that $Ri_g$ serves as a robust global parameter.

Moreover, apart from geometric asymmetry, strong shear inhomogeneity can have a significant impact on the sheared ACRBC of small radius ratios. In instability analysis, the inhomogeneity of shear leads to the outward displacement of perturbations in critical modes at high Rayleigh numbers. Meanwhile, it also causes the well-mixed convection region to shift outward under strong shear, which is reflected by the asymmetric temperature profiles in the numerical simulations.

By exploring the effect of the radius ratio on the sheared ACRBC system, we successfully match the stabilization mechanism of sheared ACRBC to that of wall-sheared RBC and answer the question that why a stable regime appears in the former. Shear inhibits the streamwise perturbations and stabilizes the thermal convection, but the asymmetry of the system and the inhomogeneity of the shear distribution can also have an important effect on flow characteristics and stability. As the thermal flow in this study is still in the classical regime, extending the current investigations to the ultimate regime poses an ongoing challenge. Does the interaction of shear and buoyancy change under very strong convection? Despite being limited by the huge demand for computational resources, this is an interesting question that deserves future exploration.

%Comparing the sheared ACRBC to the wall-sheared RBC, we are keenly aware of the inhomogeneity of the shear strength distribution due to the geometric asymmetry. Comparing how the local Richardson number varies with the normalized radius $\hat{r}$ for marginal states with different radius ratios, we find that the $Ri(\hat{r})$ curves of different radius ratios intersect at $\hat{r}^*\approx0.45$. With the new global Richardson number chosen as $Ri_g=Ri(\hat{r}^*)$, all the marginal-state curves are collapsed together surprisingly in the parameter domain $(Ra, Ri_g)$. Meanwhile, the results are also consistent with the marginal-state curve $Ra_c(Ri)$ of the stability analysis of the wall-sheared RBC in the streamwise direction, where the $Ri$ for wall-sheared RBC is clearly defined as $Ri=Ra/(PrRe_w^2)$. This demonstrates that the stabilization mechanism of sheared ACRBC is identical to that of wall-sheared RBC in the streamwise direction; the difference between sheared ACRBC and wall-sheared RBC in three dimensions arises from whether the perturbations in the spanwise direction are confined or not.
%the critical Rayleigh number $Ra_c$ varies with the rotating angular speed difference $\Omega$ in a similar trend, while a $Ra_c$ grows with an increasing growth rate as $\Omega$ increases, and finally when $Ra_c\ge10^5$, the marginal-state curve $Ra_c(\Omega)$ prominently inclines, nearly reaching a vertical orientation. 

\appendix

\section{Numerical details}
The parameters of the main simulations considered in this work are listed in the table \ref{tab:1}. The columns from left to right indicate the Rayleigh number $Ra$, the radius ratio $\eta$, the non-dimensional rotating velocity difference $\Omega$, the global Richardson number $Ri_g$, the resolution in the radial and azimuthal direction $(N_r, N_\varphi)$, the Nusselt number of heat transfer $Nu_h$ and its relative difference of two halves $\epsilon_{Nu_h}$, the Nusselt number of momentum transfer $Nu_\omega$ and its relative difference of two halves $\epsilon_{Nu_\omega}$ and the posterior check on the maximum grid spacing $\Delta_g$ by the Kolmogorov scale $\eta_K$ and the Batchelor scale $\eta_B$. The Kolmogorov scale is estimated by the global criterion $\eta_K=(\nu^3/\varepsilon)^{1/4}$, where $\varepsilon$ is the mean energy dissipation rate calculated by the equation (\ref{NuEqu}). The statistic errors are estimated by the differences between the first half and the second half, as: $\epsilon_{Nu_{h,\omega}}=|(\langle Nu_{h,\omega}\rangle_{0-T/2}-\langle Nu_{h,\omega}\rangle_{T/2-T})/(Nu-1)|$.
%\begin{table}
%\def~{\hphantom{0}}
%\begin{center}
%\resizebox{\textwidth}{!}{%
%\begin{tabular}{ccccccccccc}

\begin{longtable}{c|cccccccccccc}
\caption{Simulation parameters.}\label{tab:1}\\
\hline
$No.$ &$Ra$ & $\eta $ & $ \Omega $ &$Ri_g$ & $ N_r  $ & $ N_\varphi   $ & $ Nu_h   $ & $ \epsilon_{Nu_h} $ & $ Nu_\omega    $ & $ \epsilon_{Nu_\omega}   $ & $ \Delta_g/\eta_K $ & $ \Delta_g/\eta_B $\\
\hline
\endhead
\hline
\endfoot
%\multicolumn{11}{c}%
$1  $ & $ 10^6 $ & $ 0.3 $ & $ 0     $ & $ \infty   $ & $ 128 $ & $ 1024 $ & $ 6.410  $ & $ 0.05\%  $ & $ - $ & $ -   $ & $ 0.24 $ & $ 0.50 $\\
$2  $ & $ 10^6 $ & $ 0.3 $ & $ 0.1   $ & $ 81.76 $ & $ 128 $ & $ 1024 $ & $ 6.501  $ & $ 0.18\%  $ & $ -0.667 $ & $ 0.42\%  $ & $ 0.24 $ & $ 0.50 $\\
$3  $ & $ 10^6 $ & $ 0.3 $ & $ 0.2   $ & $ 20.44 $ & $ 128 $ & $ 1024 $ & $ 5.949  $ & $ 0.05\%  $ & $ -0.174 $ & $ 0.19\%  $ & $ 0.23 $ & $ 0.48 $\\
$4  $ & $ 10^6 $ & $ 0.3 $ & $ 0.286 $ & $ 10.00 $ & $ 128 $ & $ 1024 $ & $ 5.808  $ & $ 0.80\%  $ & $ 0.202  $ & $ 1.20\%  $ & $ 0.23 $ & $ 0.47 $\\
$5  $ & $ 10^6 $ & $ 0.3 $ & $ 0.4   $ & $ 5.11  $ & $ 128 $ & $ 1024 $ & $ 5.312  $ & $ 0.53\%  $ & $ 0.582  $ & $ 1.47\%  $ & $ 0.22 $ & $ 0.46 $\\
$6  $ & $ 10^6 $ & $ 0.3 $ & $ 0.404 $ & $ 5.01  $ & $ 128 $ & $ 1024 $ & $ 5.303  $ & $ 1.17\%  $ & $ 0.569  $ & $ 1.71\%  $ & $ 0.22 $ & $ 0.46 $\\
$7  $ & $ 10^6 $ & $ 0.3 $ & $ 0.6   $ & $ 2.27  $ & $ 128 $ & $ 1024 $ & $ 2.876  $ & $ 1.21\%  $ & $ 0.812  $ & $ 0.45\%  $ & $ 0.17 $ & $ 0.35 $\\
$8  $ & $ 10^6 $ & $ 0.3 $ & $ 0.639 $ & $ 2.00  $ & $ 128 $ & $ 1024 $ & $ 2.521  $ & $ 0.65\%  $ & $ 0.859  $ & $ 1.28\%  $ & $ 0.16 $ & $ 0.33 $\\
$9  $ & $ 10^6 $ & $ 0.3 $ & $ 0.8   $ & $ 1.28  $ & $ 128 $ & $ 1024 $ & $ 1.823  $ & $ 0.55\%  $ & $ 0.938  $ & $ 1.39\%  $ & $ 0.13 $ & $ 0.28 $\\
$10 $ & $ 10^6 $ & $ 0.3 $ & $ 0.904 $ & $ 1.00  $ & $ 128 $ & $ 1024 $ & $ 1.527  $ & $ 1.52\%  $ & $ 0.969  $ & $ 3.89\%  $ & $ 0.12 $ & $ 0.25 $\\
$11 $ & $ 10^6 $ & $ 0.3 $ & $ 1.0   $ & $ 0.82  $ & $ 128 $ & $ 1024 $ & $ 1.318  $ & $ 0.46\%  $ & $ 0.983  $ & $ 1.95\%  $ & $ 0.10 $ & $ 0.21 $\\
$12 $ & $ 10^6 $ & $ 0.3 $ & $ 1.167 $ & $ 0.60  $ & $ 129 $ & $ 1025 $ & $ 1.083  $ & $ 3.10\%  $ & $ 0.997  $ & $ 1.80\%  $ & $ 0.07 $ & $ 0.15 $\\
$13 $ & $ 10^6 $ & $ 0.3 $ & $ 1.2   $ & $ 0.57  $ & $ 128 $ & $ 1024 $ & $ 1.064  $ & $ 3.30\%  $ & $ 0.998  $ & $ 0.44\%  $ & $ 0.07 $ & $ 0.14 $\\
$14 $ & $ 10^6 $ & $ 0.5 $ & $ 0     $ & $ \infty    $ & $ 128 $ & $ 1536 $ & $ 7.286  $ & $ 0.47\%  $ & $ -   $ & $ -   $ & $ 0.26 $ & $ 0.53 $\\
$15 $ & $ 10^6 $ & $ 0.5 $ & $ 0.05  $ & $ 59.79 $ & $ 128 $ & $ 1536 $ & $ 7.008  $ & $ 0.05\%  $ & $ -1.083 $ & $ 1.97\%  $ & $ 0.25 $ & $ 0.52 $\\
$16 $ & $ 10^6 $ & $ 0.5 $ & $ 0.1   $ & $ 14.95 $ & $ 128 $ & $ 1536 $ & $ 6.775  $ & $ 0.79\%  $ & $ -0.390 $ & $ 1.15\%  $ & $ 0.25 $ & $ 0.51 $\\
$17 $ & $ 10^6 $ & $ 0.5 $ & $ 0.2   $ & $ 3.74  $ & $ 128 $ & $ 1536 $ & $ 6.285  $ & $ 0.64\%  $ & $ 0.195  $ & $ 1.92\%  $ & $ 0.23 $ & $ 0.48 $\\
$18 $ & $ 10^6 $ & $ 0.5 $ & $ 0.3   $ & $ 1.66  $ & $ 128 $ & $ 1536 $ & $ 5.252  $ & $ 2.00\%  $ & $ 0.556  $ & $ 3.72\%  $ & $ 0.21 $ & $ 0.44 $\\
$19 $ & $ 10^6 $ & $ 0.5 $ & $ 0.4   $ & $ 0.93  $ & $ 128 $ & $ 1536 $ & $ 2.077  $ & $ 1.27\%  $ & $ 0.915  $ & $ 2.66\%  $ & $ 0.14 $ & $ 0.30 $\\
$20 $ & $ 10^6 $ & $ 0.5 $ & $ 0.5   $ & $ 0.60  $ & $ 128 $ & $ 1536 $ & $ 1.257  $ & $ 3.22\%  $ & $ 0.986  $ & $ 4.10\%  $ & $ 0.10 $ & $ 0.21 $\\
$21 $ & $ 10^6 $ & $ 0.7 $ & $ 0     $ & $ \infty   $ & $ 128 $ & $ 2560 $ & $ 7.453  $ & $ 0.20\%  $ & $ -   $ & $ -   $ & $ 0.26 $ & $ 0.54 $\\
$22 $ & $ 10^6 $ & $ 0.7 $ & $ 0.02  $ & $ 81.42 $ & $ 128 $ & $ 2560 $ & $ 7.481  $ & $ 0.64\%  $ & $ -0.753 $ & $ 1.15\%  $ & $ 0.26 $ & $ 0.54 $\\
$23 $ & $ 10^6 $ & $ 0.7 $ & $ 0.03  $ & $ 36.19 $ & $ 128 $ & $ 2560 $ & $ 7.368  $ & $ 0.05\%  $ & $ -0.551 $ & $ 0.06\%  $ & $ 0.26 $ & $ 0.53 $\\
$24 $ & $ 10^6 $ & $ 0.7 $ & $ 0.05  $ & $ 13.03 $ & $ 128 $ & $ 2560 $ & $ 7.142  $ & $ 0.88\%  $ & $ -0.274 $ & $ 0.69\%  $ & $ 0.25 $ & $ 0.52 $\\
$25 $ & $ 10^6 $ & $ 0.7 $ & $ 0.057 $ & $ 10.02 $ & $ 128 $ & $ 2560 $ & $ 7.022  $ & $ 0.77\%  $ & $ -0.243 $ & $ 1.38\%  $ & $ 0.25 $ & $ 0.52 $\\
$26 $ & $ 10^6 $ & $ 0.7 $ & $ 0.081 $ & $ 4.96  $ & $ 128 $ & $ 2560 $ & $ 6.779  $ & $ 1.05\%  $ & $ -0.024 $ & $ 1.56\%  $ & $ 0.24 $ & $ 0.50 $\\
$27 $ & $ 10^6 $ & $ 0.7 $ & $ 0.1   $ & $ 3.26  $ & $ 128 $ & $ 2560 $ & $ 6.662  $ & $ 0.91\%  $ & $ 0.131  $ & $ 1.42\%  $ & $ 0.24 $ & $ 0.49 $\\
$28 $ & $ 10^6 $ & $ 0.7 $ & $ 0.128 $ & $ 1.99  $ & $ 128 $ & $ 2560 $ & $ 6.332  $ & $ 0.53\%  $ & $ 0.342  $ & $ 0.29\%  $ & $ 0.23 $ & $ 0.48 $\\
$29 $ & $ 10^6 $ & $ 0.7 $ & $ 0.15  $ & $ 1.45  $ & $ 128 $ & $ 2560 $ & $ 6.142  $ & $ 1.72\%  $ & $ 0.464  $ & $ 1.11\%  $ & $ 0.22 $ & $ 0.46 $\\
$30 $ & $ 10^6 $ & $ 0.7 $ & $ 0.17  $ & $ 1.13  $ & $ 128 $ & $ 2560 $ & $ 5.371  $ & $ 0.54\%  $ & $ 0.598  $ & $ 1.00\%  $ & $ 0.21 $ & $ 0.44 $\\
$31 $ & $ 10^6 $ & $ 0.7 $ & $ 0.181 $ & $ 0.99  $ & $ 128 $ & $ 2560 $ & $ 4.731  $ & $ 1.41\%  $ & $ 0.675  $ & $ 1.10\%  $ & $ 0.20 $ & $ 0.42 $\\
$32 $ & $ 10^6 $ & $ 0.7 $ & $ 0.2   $ & $ 0.81  $ & $ 128 $ & $ 2560 $ & $ 2.171  $ & $ 1.52\%  $ & $ 0.905  $ & $ 1.53\%  $ & $ 0.15 $ & $ 0.31 $\\
$33 $ & $ 10^6 $ & $ 0.7 $ & $ 0.23  $ & $ 0.62  $ & $ 128 $ & $ 2560 $ & $ 1.413  $ & $ 1.48\%  $ & $ 0.978  $ & $ 5.44\%  $ & $ 0.12 $ & $ 0.24 $\\
$34 $ & $ 10^6 $ & $ 0.7 $ & $ 0.233 $ & $ 0.60  $ & $ 128 $ & $ 2560 $ & $ 1.320  $ & $ 5.98\%  $ & $ 0.982  $ & $ 2.46\%  $ & $ 0.11 $ & $ 0.22 $\\
$35 $ & $ 10^6 $ & $ 0.9 $ & $ 0     $ & $ \infty   $ & $ 128 $ & $ 7680 $ & $ 7.769  $ & $ 0.07\%  $ & $ -  $ & $ -  $ & $ 0.26 $ & $ 0.55 $\\
$36 $ & $ 10^6 $ & $ 0.9 $ & $ 0.007 $ & $ 55.92 $ & $ 128 $ & $ 7680 $ & $ 7.569  $ & $ 0.05\%  $ & $ -0.324 $ & $ 0.02\%  $ & $ 0.26 $ & $ 0.54 $\\
$37 $ & $ 10^6 $ & $ 0.9 $ & $ 0.01  $ & $ 27.40 $ & $ 128 $ & $ 7680 $ & $ 7.354  $ & $ 0.27\%  $ & $ -0.254 $ & $ 0.47\%  $ & $ 0.26 $ & $ 0.53 $\\
$38 $ & $ 10^6 $ & $ 0.9 $ & $ 0.02  $ & $ 6.85  $ & $ 128 $ & $ 7680 $ & $ 6.997  $ & $ 0.37\%  $ & $ -0.060 $ & $ 1.09\%  $ & $ 0.25 $ & $ 0.51 $\\
$39 $ & $ 10^6 $ & $ 0.9 $ & $ 0.03  $ & $ 3.04  $ & $ 128 $ & $ 7680 $ & $ 6.749  $ & $ 0.82\%  $ & $ 0.157  $ & $ 0.27\%  $ & $ 0.24 $ & $ 0.50 $\\
$40 $ & $ 10^6 $ & $ 0.9 $ & $ 0.05  $ & $ 1.10  $ & $ 128 $ & $ 7680 $ & $ 5.856  $ & $ 0.04\%  $ & $ 0.562  $ & $ 0.51\%  $ & $ 0.22 $ & $ 0.45 $\\
$41 $ & $ 10^6 $ & $ 0.9 $ & $ 0.07  $ & $ 0.56  $ & $ 128 $ & $ 7680 $ & $ 3.557  $ & $ 0.04\%  $ & $ 0.837  $ & $ 0.01\%  $ & $ 0.17 $ & $ 0.36 $\\
 $  $ & $          $ & $     $ & $       $ & $       $ & $     $ & $      $ & $        $ & $         $ & $        $ & $         $ & $      $ & $      $\\
$42 $ & $ 10^7 $ & $ 0.3 $ & $ 0     $ & $ \infty  $ & $ 128 $ & $ 1024 $ & $ 11.644 $ & $ 0.79\%  $ & $ - $ & $ -   $ & $ 0.51 $ & $ 1.06 $\\
$43 $ & $ 10^7 $ & $ 0.3 $ & $ 0.1   $ & $ 81.76 $ & $ 128 $ & $ 1024 $ & $ 10.474 $ & $ 0.71\%  $ & $ -2.755 $ & $ 1.54\%  $ & $ 0.49 $ & $ 1.02 $\\
$44 $ & $ 10^7 $ & $ 0.3 $ & $ 0.2   $ & $ 20.44 $ & $ 128 $ & $ 1024 $ & $ 10.523 $ & $ 0.68\%  $ & $ -0.843 $ & $ 1.55\%  $ & $ 0.49 $ & $ 1.01 $\\
$45 $ & $ 10^7 $ & $ 0.3 $ & $ 0.4   $ & $ 5.11  $ & $ 128 $ & $ 1024 $ & $ 9.260  $ & $ 1.49\%  $ & $ 0.114  $ & $ 2.59\%  $ & $ 0.46 $ & $ 0.95 $\\
$46 $ & $ 10^7 $ & $ 0.3 $ & $ 0.6   $ & $ 2.27  $ & $ 128 $ & $ 1024 $ & $ 7.078  $ & $ 0.26\%  $ & $ 0.390  $ & $ 1.11\%  $ & $ 0.41 $ & $ 0.84 $\\
$47 $ & $ 10^7 $ & $ 0.3 $ & $ 0.8   $ & $ 1.28  $ & $ 128 $ & $ 1024 $ & $ 2.209  $ & $ 0.23\%  $ & $ 0.895  $ & $ 6.46\%  $ & $ 0.25 $ & $ 0.53 $\\
$48 $ & $ 10^7 $ & $ 0.3 $ & $ 1     $ & $ 0.82  $ & $ 128 $ & $ 1024 $ & $ 1.561  $ & $ 2.94\%  $ & $ 0.961  $ & $ 6.70\%  $ & $ 0.20 $ & $ 0.41 $\\
$49 $ & $ 10^7 $ & $ 0.3 $ & $ 1.2   $ & $ 0.57  $ & $ 128 $ & $ 1024 $ & $ 1.279  $ & $ 4.30\%  $ & $ 0.989  $ & $ 13.86\% $ & $ 0.18 $ & $ 0.36 $\\
$50 $ & $ 10^7 $ & $ 0.5 $ & $ 0     $ & $ \infty   $ & $ 128 $ & $ 1536 $ & $ 13.288 $ & $ 0.36\%  $ & $ - $ & $ - $ & $ 0.54 $ & $ 1.12 $\\
$51 $ & $ 10^7 $ & $ 0.5 $ & $ 0.05  $ & $ 59.79 $ & $ 128 $ & $ 1536 $ & $ 13.512 $ & $ 0.43\%  $ & $ -2.555 $ & $ 0.62\%  $ & $ 0.54 $ & $ 1.12 $\\
$52 $ & $ 10^7 $ & $ 0.5 $ & $ 0.1   $ & $ 14.95 $ & $ 128 $ & $ 1536 $ & $ 12.422 $ & $ 0.57\%  $ & $ -1.552 $ & $ 0.03\%  $ & $ 0.52 $ & $ 1.08 $\\
$53 $ & $ 10^7 $ & $ 0.5 $ & $ 0.2   $ & $ 3.74  $ & $ 128 $ & $ 1536 $ & $ 12.199 $ & $ 1.55\%  $ & $ -0.470 $ & $ 2.42\%  $ & $ 0.50 $ & $ 1.04 $\\
$54 $ & $ 10^7 $ & $ 0.5 $ & $ 0.3   $ & $ 1.66  $ & $ 128 $ & $ 1536 $ & $ 10.330 $ & $ 1.01\%  $ & $ 0.348  $ & $ 1.37\%  $ & $ 0.48 $ & $ 0.99 $\\
$55 $ & $ 10^7 $ & $ 0.5 $ & $ 0.4   $ & $ 0.93  $ & $ 128 $ & $ 1536 $ & $ 7.515  $ & $ 1.09\%  $ & $ 0.503  $ & $ 3.15\%  $ & $ 0.41 $ & $ 0.84 $\\
$56 $ & $ 10^7 $ & $ 0.5 $ & $ 0.45  $ & $ 0.74  $ & $ 128 $ & $ 1536 $ & $ 3.119  $ & $ 0.74\%  $ & $ 0.841  $ & $ 0.86\%  $ & $ 0.29 $ & $ 0.61 $\\
$57 $ & $ 10^7 $ & $ 0.5 $ & $ 0.5   $ & $ 0.60  $ & $ 128 $ & $ 1536 $ & $ 1.452  $ & $ 12.70\% $ & $ 0.969  $ & $ 14.48\% $ & $ 0.19 $ & $ 0.40 $\\
$58 $ & $ 10^7 $ & $ 0.7 $ & $ 0     $ & $ \infty   $ & $ 128 $ & $ 2560 $ & $ 13.971 $ & $ 0.14\%  $ & $ -  $ & $ -  $ & $ 0.55 $ & $ 1.14 $\\
$59 $ & $ 10^7 $ & $ 0.7 $ & $ 0.02  $ & $ 81.42 $ & $ 128 $ & $ 2560 $ & $ 14.402 $ & $ 0.14\%  $ & $ -2.399 $ & $ 1.58\%  $ & $ 0.55 $ & $ 1.15 $\\
$60 $ & $ 10^7 $ & $ 0.7 $ & $ 0.05  $ & $ 13.03 $ & $ 128 $ & $ 2560 $ & $ 13.499 $ & $ 0.14\%  $ & $ -1.613 $ & $ 1.18\%  $ & $ 0.53 $ & $ 1.11 $\\
$61 $ & $ 10^7 $ & $ 0.7 $ & $ 0.1   $ & $ 3.26  $ & $ 128 $ & $ 2560 $ & $ 12.110 $ & $ 0.03\%  $ & $ -0.235 $ & $ 7.07\%  $ & $ 0.51 $ & $ 1.05 $\\
$62 $ & $ 10^7 $ & $ 0.7 $ & $ 0.15  $ & $ 1.45  $ & $ 128 $ & $ 2560 $ & $ 13.525 $ & $ 0.76\%  $ & $ -0.452 $ & $ 1.20\%  $ & $ 0.49 $ & $ 1.02 $\\
$63 $ & $ 10^7 $ & $ 0.7 $ & $ 0.2   $ & $ 0.81  $ & $ 128 $ & $ 2560 $ & $ 9.362  $ & $ 0.39\%  $ & $ 0.381  $ & $ 1.80\%  $ & $ 0.44 $ & $ 0.90 $\\
$64 $ & $ 10^7 $ & $ 0.7 $ & $ 0.23  $ & $ 0.62  $ & $ 128 $ & $ 2560 $ & $ 5.555  $ & $ 1.31\%  $ & $ 0.665  $ & $ 1.16\%  $ & $ 0.35 $ & $ 0.73 $\\
$65 $ & $ 10^7 $ & $ 0.9 $ & $ 0     $ & $ \infty  $ & $ 128 $ & $ 7680 $ & $ 14.640 $ & $ 0.16\%  $ & $ - $ & $ - $ & $ 0.56 $ & $ 1.16 $\\
$66 $ & $ 10^7 $ & $ 0.9 $ & $ 0.007 $ & $ 55.92 $ & $ 128 $ & $ 7680 $ & $ 14.372 $ & $ 0.14\%  $ & $ -1.696 $ & $ 0.81\%  $ & $ 0.55 $ & $ 1.15 $\\
$67 $ & $ 10^7 $ & $ 0.9 $ & $ 0.01  $ & $ 27.40 $ & $ 128 $ & $ 7680 $ & $ 14.715 $ & $ 0.02\%  $ & $ -1.357 $ & $ 0.84\%  $ & $ 0.55 $ & $ 1.15 $\\
$68 $ & $ 10^7 $ & $ 0.9 $ & $ 0.017 $ & $ 9.48  $ & $ 128 $ & $ 7680 $ & $ 13.922 $ & $ 0.48\%  $ & $ -1.183 $ & $ 0.75\%  $ & $ 0.54 $ & $ 1.12 $\\
$69 $ & $ 10^7 $ & $ 0.9 $ & $ 0.03  $ & $ 3.04  $ & $ 128 $ & $ 7680 $ & $ 12.234 $ & $ 0.47\%  $ & $ -0.109 $ & $ 1.01\%  $ & $ 0.51 $ & $ 1.06 $\\
$70 $ & $ 10^7 $ & $ 0.9 $ & $ 0.05  $ & $ 1.10  $ & $ 128 $ & $ 7680 $ & $ 11.377 $ & $ 0.76\%  $ & $ 0.360  $ & $ 3.90\%  $ & $ 0.49 $ & $ 1.01 $\\
$71 $ & $ 10^7 $ & $ 0.9 $ & $ 0.07  $ & $ 0.56  $ & $ 128 $ & $ 7680 $ & $ 6.450  $ & $ 0.46\%  $ & $ 0.675  $ & $ 0.09\%  $ & $ 0.38 $ & $ 0.79 $\\
%\end{tabular}%

%\end{center}
\end{longtable}

\backsection [Acknowledgements] {}

\backsection [Funding]{This work was supported by the National Natural Science Foundation of China under grant no. 11988102, and the New Cornerstone Science Foundation through the New Cornerstone Investigator Program and the XPLORER PRIZE.}

\backsection[Declaration of interests] {The authors report no conflict of interest.}

%\bibliographystyle{jfm}
%\bibliography{jfm}
%Use of the above commands will create a bibliography using the .bib file. Shown below is a bibliography built from individual items.

\bibliographystyle{jfm}
\bibliography{main}

\end{document}